\documentclass[journal]{aa}

\usepackage{graphicx}
\usepackage{amsmath}
\usepackage{amssymb}
\usepackage{verbatim}
\usepackage{array}
\usepackage{txfonts}
\usepackage{natbib}
\usepackage[switch,pagewise]{lineno}
\usepackage{multirow}
\usepackage{float}

\bibpunct{(}{)}{;}{a}{}{,}

\newcommand{\celltspace}{\rule{0pt}{2.8ex}}
\newcommand{\cellbspace}{\rule[-1.4ex]{0pt}{0pt}}

\newcommand{\kms}{\,km\,s$^{-1}$}

\newcommand{\dunit}{\,cm$^{-3}$}

\newcommand{\lsol}{\,L$_{\odot}$}
\newcommand{\wunit}{\,M$_{\odot}$\,yr$^{-1}$}
\newcommand{\kev}{\,keV}
\newcommand{\mev}{\,MeV}
\newcommand{\gev}{\,GeV}
\newcommand{\tev}{\,TeV}
\newcommand{\pev}{\,PeV}
\newcommand{\gray}{$\gamma$-ray}
\newcommand{\grays}{$\gamma$-rays}

\begin{document}

\title{Interstellar gamma-ray emission from cosmic rays \\in star-forming galaxies}
\titlerunning{Interstellar gamma-ray emission from star-forming galaxies}
\author{P.~Martin}
\authorrunning{Martin~P.}
\institute{Institut de Recherche en Astrophysique et Plan\'etologie, UPS/CNRS, UMR5277, 31028 Toulouse cedex 4, France}
\date{Received: 23 Dec 2013 / Accepted: 8 Jan 2014}
\abstract{{\em Fermi}/LAT observations of star-forming galaxies in the $\sim$0.1-100\gev\ range have made possible a first population study. Evidence was found for a correlation between \gray\ luminosity and tracers of the star formation activity. Studying galactic cosmic rays (CRs) in various global conditions can yield information about their origin and transport in the interstellar medium (ISM).}{This work addresses the question of the scaling laws that can be expected for the interstellar \gray\ emission as a function of global galactic properties, with the goal of establishing whether the current experimental data in the GeV range can be constraining.}{I developed a 2D model for the non-thermal emissions from steady-state CR populations interacting with the ISM in star-forming galaxies. Most CR-related parameters were taken from Milky Way studies, and a large number of galaxies were then simulated with sizes from 4 to 40\,kpc, several gas distributions, and star formation rates (SFR) covering six orders of magnitude.}{The evolution of the \gray\ luminosity over the 100\kev-100\tev\ range is presented, with emphasis on the contribution of the different emission processes and particle populations, and on the transition between transport regimes. The model can reproduce the normalisation and trend inferred from the \textit{Fermi}/LAT population study over most of the SFR range. This is obtained with a plain diffusion scheme, a single diffusion coefficient, and the assumption that CRs experience large-scale volume-averaged interstellar conditions. There is, however, no universal relation between high-energy \gray\ luminosity and star formation activity, as illustrated by the scatter introduced by different galactic global properties and the downturn in \gray\ emission at the low end.}{The current \textit{Fermi}/LAT population study does not call for major modifications of the transport scheme for CRs in the Milky Way when extrapolated to other systems, probably because the uncertainties are still too large. Additional constraints may be expected from doubling the {\em Fermi}/LAT exposure time and later from observing at TeV energies with the \textit{Cherenkov Telescope Array}.}
\keywords{Acceleration of particles -- Cosmic rays -- Gamma rays: galaxies}
\maketitle

\section{Introduction}
\label{intro}

\indent Star-forming galaxies harbour large-scale populations of energetic particles accelerated in violent phenomena associated with stellar evolution, such as supernova explosions, colliding-wind binaries, or pulsar wind nebulae. These galactic cosmic rays (hereafter CRs) interact with the various components of the interstellar medium (ISM): gas, turbulence, radiation, and magnetic fields. These interactions determine the transport of CRs from sources to intergalactic medium, and are responsible for their confinement and accumulation in the galactic volume over a large number of acceleration episodes (of the order of $\sim10^5$ for the Milky Way). The existence of such large-scale populations is attested by various extended emissions, notably in radio through synchrotron emission, but also in \grays\ through inverse-Compton scattering, Bremsstrahlung, and inelastic nuclear collisions. These emissions are signatures of the combined processes of CR acceleration and transport, and their interpretation in the frame of other astronomical data can provide us with a better picture of the life cycle of CRs \citep[see][for a review]{Strong:2007}.

\indent Understanding CRs is relevant to many fields beyond high-energy astrophysics. This is mostly because they are more than a simple side effect of the end stages of stellar evolution. Galactic cosmic rays interact with the ISM and can significantly modify the environment in which they are flowing. They can heat and ionise the gas, especially at the centre of dense molecular clouds, thereby impacting star formation conditions \citep{Papadopoulos:2010}; through their diffusion, CRs can alter the turbulence of the interstellar medium, with consequences on their own transport \citep{Ptuskin:2006}; they can contribute to large-scale outflows, an important feedback effect on star formation over cosmic times \citep{Hanasz:2013}; CRs may play a role in shaping large-scale galactic magnetic fields through a dynamo process \citep{Hanasz:2004}; and last, CRs from classical astrophysical objects can hide the signatures of more exotic components of the Universe such as dark matter \citep{Delahaye:2010}. For all these reasons, understanding the non-thermal content and emission of star-forming galaxies is crucial.

\indent The current generation of high-energy (HE) and very-high-energy (VHE) \gray\ telescopes has enabled us to detect several external galaxies that shine at photon energies $>100$\mev\ as a result of CRs interacting with their ISM (and not because of a central black hole activity, as is the case for the vast majority of detected \gray-emitting external galaxies). The {\em Fermi}/LAT space-borne pair-creation telescope has permitted the detection of five external galaxies at GeV energies (SMC, LMC, M31, M82, and NGC253), with two other systems being possible candidates \citep[NGC4945 and NGC1068, whose emission may be contaminated by central black hole activity; see][]{Abdo:2009l,Abdo:2010d,Abdo:2010e,Abdo:2010f,Ackermann:2012}. In the same time, the HESS and VERITAS ground-based Cherenkov telescopes have permitted the detection at TeV energies of NGC253 and M82, respectively \citep{Acero:2009,Abramowski:2012,Acciari:2009}. Star-forming galaxies can now be considered as an emerging class of \gray\ sources, with a notable increase of the number of detected objects from the previous generation of instruments \citep[only the LMC was detected by CGRO/EGRET; see][]{Sreekumar:1992}.

\indent Even with such a limited sample, it is tempting to perform a population study to find clues about the main parameters regulating CR populations. From the {\em Fermi}/LAT sample of five objects plus an estimate for the global emission of the Milky Way and an upper limit for M33, the work of \citet{Abdo:2010f} showed first hints for a slightly non-linear correlation between \gray\ luminosity and star formation rate (SFR). A more systematic study by \citet{Ackermann:2012}, including upper limits for about sixty luminous and ultra-luminous infrared galaxies, found further evidence for a quasi-linear correlation between \gray\ luminosity and tracers of the star formation activity. Such a relationship can be expected to first order, because the production of CRs is expected to scale with the rate of massive star formation; but it is surprising that the correlation is so close to linear, over such a wide range of galactic properties, from dwarfs like the SMC to interacting systems like Arp 220. 

\indent The interpretation of this correlation therefore raised the question of the scaling laws that can be expected for the diffuse \gray\ emission as a function of global galactic properties. Clarifying this at the theoretical level is required to establish whether the current experimental data can be constraining in any way. This is all the more needed because the sample of detected objects is small, the coverage over the GeV-TeV range is uneven, and the spectral characterization is still rather limited. In addition, the determination of global properties of external systems, such as total mass or star formation rate, is in itself a challenge, and there is most likely some intrinsic scatter in the actual physical correlations.

\indent That star-forming galaxies might become a class of \gray\ sources was anticipated early, well before the launch of the \textit{Fermi} satellite \citep[see for instance][]{Volk:1996, Pavlidou:2001,Torres:2004}. The interest in the subject extended beyond the study of nearby objects, because this emission cumulated over cosmic time might account for a significant fraction of the isotropic diffuse \gray\ background \citep{Pavlidou:2002,Thompson:2007, Ackermann:2012}. Before and after the launch of the \textit{Fermi} satellite, the \gray\ emission from individual starbursts was studied by several authors, with the focus being mainly on M82 and NGC253 \citep{Persic:2008,deCeadelPozo:2009,Lacki:2011,Paglione:2012}.

\indent The purpose of this paper is to provide a more global picture of the evolution of the \gray\ interstellar emission from star-forming galaxies, in the context of the recent population study conducted on the basis of {\em Fermi}/LAT observations. Instead of trying to fit a model to a particular object, I defined a global framework by fixing as many cosmic-ray-related parameters as possible from Milky Way studies, leaving only a few global quantities as independent variables. Here, I present a generic model for the transport of CRs and the associated non-thermal interstellar emissions in star-forming galaxies. Simulating systems with various sizes and gas distributions thus allows investigating the impact of global galactic parameters on the \gray\ output. In a first part, I introduce the different components and related assumptions of the model, together with the set of synthetic galaxies that was considered. Then, the \gray\ emission over 100\kev-100\tev\ for this series of synthetic galaxies is presented, and its evolution with global properties is analysed. In a subsequent part, the scaling of \gray\ luminosity with global galactic properties is derived for soft, high-energy, and very-high-energy \gray\ ranges, and is compared with observations in the case of the high-energy band. Last, the impact of the model parameters is discussed, and the connection to the far-infrared - radio correlation is addressed.

\section{Model}
\label{model}

\indent The model is based on a modified version of the GALPROP public code\footnote{GALPROP is a software package for modelling the propagation of CRs in the Galaxy and their interactions in the ISM. It allows simultaneous and coherent predictions of the CR population in the interplanetary and interstellar medium, gamma-ray emission, and recently also synchrotron radiation intensity and polarization. See http://galprop.stanford.edu/ and http://sourceforge.net/projects/galprop/ for the latest developments.}, originally designed for the Milky Way and adapted here to model any star-forming galaxy. For the most part, this was achieved by implementing general models for the interstellar medium components, as described below. This generic model uses as input and for calibration some characteristics of a typical GALPROP modelling of the Milky Way in the plain diffusion case: Model z04LMPDS in Table 1 of \citet{Strong:2010}, referred to in the following as the \textit{reference GALPROP run}.

\indent All the synthetic galaxies discussed in this work are two-dimensional in space with cylindrical symmetry, and are defined by a disk radius $R_{\textrm{max}}$ and a halo height $z_{\textrm{max}}$. The model is therefore better suited to simulate grand-design spiral galaxies than strongly irregular or interacting galaxies, but it can provide insights into the effects of galaxy size and star formation conditions on non-thermal emissions.

\subsection{Gas}
\label{model_gas}

\indent The interstellar gas distribution was modelled through simple analytical functions of the galactocentric radius $R$ and height above the galactic plane $z$. I considered only atomic and molecular gas, and neglected ionised gas. A ratio of He/H=0.11 was assumed for the interstellar gas. 

\indent Three types of gas distributions are discussed in the following: a disk of atomic hydrogen with exponential density profile in radius and height (three parameters: scale radius, scale height, and peak atomic gas density); a central core of molecular gas with uniform density (three parameters: core radius, core height, and molecular gas density); a ring of molecular gas with uniform density (four parameters: ring inner and outer radii, ring height, and molecular gas density). These profiles are intended to reproduce a typical star-forming galaxy, where star formation can operate at a low surface density in a disk, and/or be concentrated in more compact regions like central cores (such as in M82 and NGC253), or in spiral arms or rings (such as in the Milky Way or M31). The different models and their parameters are summarised below.

\indent An important aspect of the model, as becomes obvious in the following, is that the gas content and distribution drives many other components, such as the magnetic field model, the infrared energy density of the interstellar radiation field, or the CR source distribution. 

\subsection{Magnetic field}
\label{model_mag}

\indent The magnetic field strength is assumed to depend at each point on the vertical gas column density, according to
\begin{equation}
\label{eq_bfield}
B(R,z)=B_0 \, \left(\frac{\Sigma_{\textrm{gas}}(R)}{\Sigma_{0}} \right)^a \, e^\frac{-|z|}{z_{B}} ,
\end{equation}
where $\Sigma_{\textrm{gas}}$ is the gas surface density. Such a scaling of the mid-plane magnetic field value can be justified by different theoretical expectations, from hydrostatic equilibrium to turbulent magnetic field amplification \citep[see][]{Schleicher:2013}. The vertical profile is assumed to be exponential, and I adopted a magnetic field scale height equal to the halo size \citep[following][and quite similar to the \textit{reference GALPROP run}; the influence of this assumption is examined in Sect. \ref{res_params}]{Sun:2008}. The exponent $a$ was set to a value of 0.6 to match the observed far-infrared - radio continuum correlation, as discussed in Sect. \ref{res_radio}. The normalisation doublet $(B_0, \Sigma_{0})$ was set to (5\,$\mu$G, 4\,M$_{\odot}$\,pc$^{-2}$), from a comparison with the \textit{reference GALPROP run}, as explained in Sect. \ref{model_calib}.

\indent I did not consider a global topology model for the magnetic field, or a separation into regular and turbulent components, because the diffusion of CRs is treated in a homogeneous and isotropic way. A recent GALPROP modelling using a more complete magnetic field model can be found in \citet{Orlando:2013}. Theoretically, anisotropic diffusion along field lines can be expected, depending on the ratio of turbulent field to regular field strength, and lead for instance to channelling and clustering of particles preferentially along spiral arms \citep{Dosch:2011}. But, the reality and impact of this compared with the homogeneous and isotropic approximation is poorly constrained at the experimental level \citep[see the references and discussion in][]{Jaffe:2013}.

\subsection{Radiation field}
\label{model_isrf}

\indent The interstellar radiation field model has three main components: the cosmic microwave background (CMB), the infrared dust emission (IR), and the stellar emission (UVO). The first contribution is implemented straightforwardly as a blackbody with a temperature of $T_{\textrm{CMB}}$=2.7\,K. The other two components are dependent on the particular galactic setting and ideally result from a complex radiation transfer problem involving a variety of stellar sources and a mixture of dust grains, each with specific spatial distributions.

\indent The interstellar radiation field model (ISRF) is based on the work by \citet{Draine:2007a}. Interstellar dust is irradiated by a distribution of starlight intensities with the $I_{\textrm{M83}}(\lambda)$ spectrum determined by \citet{Mathis:1983} for the local Galactic environment and scaled by a dimensionless factor $U$. The resulting dust emission $I_{\textrm{dust}}(\lambda)$ in the 1-10000\,$\mu$m wavelength band and normalized per hydrogen atom is available for various combinations of the model parameters (see the appendix). 

\indent The infrared intensity distribution over the galactic volume ideally results from integrating at each point the contribution of all sources, possibly affected by absorption, reemission, and scattering \citep[see for instance][]{Ackermann:2012b}. As a simplification, the energy density of the infrared dust component in the ISRF model was computed as
\begin{align}
\label{eq_ir}
u_{\textrm{IR}}(R,z)= \frac{4 \pi}{c} \, E_{\textrm{dust}} \, \Sigma_{\textrm{thres}} \, \left( \frac{\Sigma_{\textrm{gas}}(R)}{\Sigma_{\textrm{thres}}} \right)^N \, e^\frac{-|z|}{z_{\textrm{IR}}} ,\\
\textrm{where }  E_{\textrm{dust}} = \int I_{\textrm{dust}}(\lambda) d\lambda ,
\end{align}
where $\Sigma_{\textrm{thres}}$ is a threshold gas surface density for star formation and $N$ the index of the Schmidt-Kennicutt relation (see the definitions below). The quantity $c$ is the speed of light. The intensity at each point $(R,z)$ in the volume is thus assumed to be dominated by dust emission from the same galactocentric radius $R$, and the energy density is taken to decrease over a scale height $z_{\textrm{IR}}$=2\,kpc above and below the galactic plane. The latter value was set by comparison with the \textit{reference GALPROP run}, as explained in Sect. \ref{model_calib}. The approximation adopted implements an IR energy density in the galactic plane that has a minimum value for the threshold of star formation, and then scales with the star formation rate according to the term in parentheses\footnote{With the adopted parameterization for the infrared dust emission, the conversion factor between star formation rate and infrared luminosity in the 8-1000\,$\mu$m band is 1.3\,10$^{-10}$\wunit/\lsol. This can be compared with 1.7\,10$^{-10}$\wunit/\lsol\ in the original formulation by \citet{Kennicutt:1998} using a Salpeter initial mass function \citep{Salpeter:1955}, or with 1.3\,10$^{-10}$\wunit/\lsol\ when using a Chabrier initial mass function \citep{Chabrier:2003}.}. I neglected the cirrus emission arising from evolved star heating, which can increase the IR energy density by up to a factor of 2 \citep{Kennicutt:2009}, but I tested the impact of varying the IR field intensity, as discussed in Sect. \ref{res_params}.

\indent The mid-plane value for the starlight energy density distribution was assumed to be the $I_{\textrm{M83}}$ field scaled by the dust-weighted mean starlight intensity $\langle U \rangle$,
\begin{align}
\label{eq_uvo}
u_{\textrm{UVO}}(R,z)= \frac{4 \pi}{c} \, \langle U \rangle \, E_{\textrm{M83}} \, e^\frac{-|z|}{z_{\textrm{UVO}}} ,\\
\textrm{where }  E_{\textrm{M83}} = \int I_{\textrm{M83}}(\lambda) d\lambda
\end{align}
With the parameters adopted (see the appendix), this mean starlight intensity is $\langle U \rangle$=3.4, so it is about 3 times stronger everywhere in the galactic disk than in the local Galactic environment. Getting the same field everywhere in the galactic disk is probably not realistic. The stellar photon field experienced by CRs is expected to be higher in regions of intense star formation, especially because this is where they are released into the ISM, but these effects occur on scales that are similar to or below the typical spatial resolution of the model. The energy density of the starlight component is taken to decrease over a scale height $z_{\textrm{UVO}}$=1\,kpc above and below the galactic plane, according to a comparison with the \textit{reference GALPROP run} (see Sect. \ref{model_calib}).

\subsection{Cosmic-ray source}
\label{model_src}

\indent The main sources of CRs are thought to be supernova remnants, possibly with some contribution of other objects such as pulsars, pulsar wind nebula, or colliding-wind binaries. Since most of these are end products of massive star evolution with lifetimes shorter than that of their progenitors, the CR source distribution was assumed by default to follow the star formation distribution. The latter is determined from the gas distribution using the empirical Schmidt-Kennicutt relation
\begin{equation}
\label{eq_sk}
\Sigma_{\textrm{SFR}}=A\,\Sigma_{\textrm{gas}}^{N} \quad \textrm{ for }  \Sigma_{\textrm{gas}} > \Sigma_{\textrm{thres}} ,
\end{equation}
where $\Sigma_{\textrm{SFR}}$ and $\Sigma_{\textrm{gas}}$ are the star formation rate and atomic+molecular gas surface density (in M$_{\odot}$\,yr$^{-1}$\,kpc$^{-2}$ and M$_{\odot}$\,pc$^{-2}$ units), and with $A=2.5\,10^{-4}$ and $N=1.4$ the canonical values \citep{Kennicutt:1998}. I used a threshold at $\Sigma_{\textrm{thres}} = $4\,M$_{\odot}$\,pc$^{-2}$, which is a lower limit from spatially resolved observations of galaxies in \citet{Kennicutt:1998}. The relation is observed to hold from galactic averages down to kpc averages. At each galactocentric radius, the vertical distribution of the star formation rate was taken to be proportional to the gas density.

\indent The total CR source luminosity was assumed by default to be proportional to the total star formation rate, using as normalisation the \textit{reference GALPROP run}. The following CR input power integrated over 100\mev-100\gev\ was used:
\begin{align}
\label{eq_crpower}
Q_{CRp,e} &= Q_{MW} \, \frac{\textrm{SFR}}{1.9 \, \textrm{M}_{\odot}\,\textrm{yr}^{-1}} , \\
\textrm{where }  Q_{MW} &= 7.10\,10^{40} \, \textrm{erg\,s}^{-1} \, \textrm{ for CRp }  \notag \\
\textrm{           }  Q_{MW} &= 1.05\,10^{39} \, \textrm{erg\,s}^{-1} \, \textrm{ for CRe }  \notag ,
\end{align}
where CRp stands for cosmic-ray hydrogen and helium nuclei, and CRe stands for cosmic-ray electrons. The SFR normalisation was obtained by integrating Eq. \ref{eq_sk} over the Galaxy given the gas distribution of the original GALPROP model; the result lies in the range of current estimates \citep{Diehl:2006}, especially if one considers that the scatter in the Schmidt-Kennicutt relation can reach an order of magnitude \citep{Kennicutt:1998}. The relation between total star formation rate and CR luminosity is probably far from straightforward because of a possible dependence of the stellar initial mass function or particle acceleration process on the interstellar conditions, but taking this into account is beyond the scope of the present work.

\indent The CR injection spectra used in this work were consistently taken from the \textit{reference GALPROP run}. Particles at source have a power-law distribution in rigidity with a break, following
\begin{align}
\label{eq_crspec}
\frac{dQ_{CRp,e}}{dR} &= Q_{0} \, \left( \frac{R}{R_{0}} \right)^{-q} , \\
\textrm{with }  q &= 1.80 \textrm{ for } R \leq R_{b}  \notag \\
\textrm{       }  q &= 2.25 \textrm{ for } R  > R_{b}  \notag \\
\textrm{       }  R_{b} &= 9\,\textrm{GV} \, \textrm{ for CRp }  \notag \\
\textrm{       }  R_{b} &= 4\,\textrm{GV} \, \textrm{ for CRe } , \notag
\end{align}
This means that the same electron-to-proton ratio at injection was used for all synthetic galaxies. This may introduce errors because acceleration can be expected to proceed differently in environments where magnetic and radiation energy densities are a factor of 100-1000 higher (as is the case in the cores of starburst galaxies). But these differences are currently poorly constrained at the experimental level, and it is interesting enough to assess the effects of global galactic properties for a given CR injection spectrum. Moreover, the assumption that particle acceleration in other galaxies works similarly to what we think takes place in the Milky Way can be tested by comparison with the emerging class of \gray\ star-forming galaxies.

\subsection{Transport}
\label{model_trans}

\indent At galactic scales, the spatial transport of CRs in the ISM is currently understood as being ruled by two main processes: advection in large-scale outflows such as galactic winds or superbubbles breaking out into the halo, and diffusion on interstellar magnetohydrodynamic turbulence \citep{Strong:2007}. While the former implies adiabatic energy losses, the latter may be associated with reacceleration of the CRs. Recent studies seem to dismiss the need for reacceleration, and show that a large set of cosmic-ray-related observations can be reasonably well accounted for from a pure diffusion scheme with isotropic and homogeneous properties \citep[][note, however, that the good fit to the data is obtained by simultaneously tuning the source and transport parameters]{Strong:2011}.  Several discrepancies between such predictions and observations remain, such as the so-called gradient problem, connected with the longitude distribution of the diffuse interstellar \gray\ emission, and the large-scale cosmic-ray anisotropy, which is observed to be smaller than expected at high energies. Several authors have shown that these can be resolved by adopting inhomogeneous and anisotropic diffusion schemes \citep{Evoli:2012}.

\indent For simplicity and as a starting point, the spatial transport of CRs in the present model of star-forming galaxy was assumed to occur by energy-dependent diffusion in a homogeneous and isotropic way. For the diffusion coefficient, I used the value from the \textit{reference GALPROP run}
\begin{align}
D_{xx}=3.4 \,10^{28} \, \beta \left( \frac{R}{4\,\textrm{GV}} \right)^{0.5} \, \textrm{cm}^{2} \, \textrm{s}^{-1} \quad \textrm{for} \quad R \geq 4\,\textrm{GV} \notag \\
D_{xx}=3.4 \,10^{28} \, \textrm{cm}^{2} \, \textrm{s}^{-1} \quad \textrm{for} \quad R < 4\,\textrm{GV} ,
\end{align}
where $R$ is the particle rigidity and $\beta=v/c$, the ratio of particle velocity to speed of light.

\indent The spatial transport of CRs in other galaxies than the Milky Way may take place in a different way. In systems harbouring a large starburst region, the interstellar turbulence may be strongly enhanced and result in different diffusion conditions, or strong outflows out of the disk may be generated and advect particles away. I discuss these two possibilities in Sect. \ref{res_params}.

\begin{table}[t!]
\caption{List of the galaxy configurations in the initial sample}
\begin{center}
\begin{tabular}{|c|c|c|c|}
\hline
\celltspace $R_{\textrm{max}}$ & $z_{\textrm{max}}$ & HI & H$_2$ \cellbspace \\
\hline
 2.0 & 1.0 & disk with $R_{\textrm{H}}$=1.0 & core with $R_{\textrm{core}}$=0.1 \\
\hline
 - & - & - & core with $R_{\textrm{core}}$=0.2 \\
\hline
 5.0 & 1.0 & disk with $R_{\textrm{H}}$=2.5 & core with $R_{\textrm{core}}$=0.2 \\
\hline
 - & - & - & core with $R_{\textrm{core}}$=0.5 \\
\hline
 - & - & - & core with $R_{\textrm{core}}$=1.0 \\
\hline
 - & - & - & ring with $(R_{i}, R_{o})=(1.5,2.0)$ \\
\hline
 - & - & - & ring with $(R_{i}, R_{o})=(2.0,3.0)$ \\
\hline
 10.0 & 2.0 & disk with $R_{\textrm{H}}$=5.0 & core with $R_{\textrm{core}}$=0.5 \\
\hline
 - & - & - & core with $R_{\textrm{core}}$=1.0 \\
\hline
 - & - & - & ring with $(R_{i}, R_{o})=(2.0,3.0)$ \\
\hline
 - & - & - & ring with $(R_{i}, R_{o})=(4.0,6.0)$ \\
\hline
 20.0 & 4.0 & disk with $R_{\textrm{H}}$=10.0 & core with $R_{\textrm{core}}$=1.0 \\
\hline
 - & - & - & ring with $(R_{i}, R_{o})=(4.0,6.0)$ \\
\hline
 - & - & - & ring with $(R_{i}, R_{o})=(8.0,12.0)$ \\
\hline
\end{tabular}
\end{center}
Note to the Table: The first two columns are the maximum radius and half-height of the galactic volume, respectively. The third column gives the scale radius of the exponential distribution of atomic gas (the vertical distribution is exponential with a scale height of 100\,pc). The fourth column indicates whether the molecular gas distribution is a uniform core or uniform ring, and gives the corresponding core radius or inner/outer radii (the vertical distribution is uniform with a half-height of 100\,pc). All lengths are in kpc units. For each model, 7 molecular gas densities were tested: $n_{\textrm{H}_2}=$5, 10, 20, 50, 100, 200, 500\,H$_2$\dunit, for the same peak atomic gas density of $n_{\textrm{H}}=$2\,H\dunit.
\label{tab_modellist}
\end{table}

\subsection{Test and calibration}
\label{model_calib}

\indent The star-forming galaxy model described above has a total of five undetermined parameters. One of these is the index of the magnetic field model, and it was set using the constraint of the far-infrared - radio correlation (see Sect. \ref{res_radio}). The other four parameters are the normalisation doublet $(B_0,\Sigma_0)$ for the magnetic field model, and the scale heights $z_{\textrm{IR}}$ and $z_{\textrm{UVO}}$ for the ISRF model. These parameters were set by a comparison with the reference GALPROP modelling of the Milky Way, which is basically what the code was optimised for. From the built-in axisymmetric atomic, molecular, and ionised gas distributions, models for the ISRF, magnetic field, and source distribution were derived using the assumptions detailed above. A typical run was then performed with these modified conditions, and its results were compared with those of the original GALPROP calculation.

\indent The comparison was first made on the ISM conditions, using source-weighted and volume averages for the energy densities of the radiation and magnetic ISM components (these averages are relevant to the conditions experienced by CRs at injection and over their long-term transport, respectively). From this, the four undefined parameters listed above were set. Applied to the Milky Way, the generic model for star-forming galaxies results in radiation (respectively magnetic) energy densities with volume and source-weighted averages that are different by about $\sim$20-30\% (respectively $\sim$5-10\%) from those of the original GALPROP setup. Then, the total Galactic non-thermal outputs were compared and were found to agree very well, without any need for further tuning of the model. The radio emission over 1\,MHz-1\,THz differs at most by 6\% at low frequencies\footnote{Note that we compare the output of two different models for the unabsorbed synchrotron emission over a very large frequency range. At the observational level, free-free absorption causes a turnover at low frequencies, and free-free and thermal dust emission dominate at high frequencies. Eventually, unabsorbed non-thermal emission dominates in the range $\sim$100\,MHz-10\,GHz.}, while the \gray\ emission over 100\kev-100\tev\ shows a maximum deviation of 5\%. When integrating the emission over $>$100\mev, the differences in total luminosity are even smaller. The similarity of the results to that of the original GALPROP calculation applies not only to the total emission but also to its components: inverse-Compton, Bremsstrahlung, and pion decay.

\subsection{Setup}
\label{model_setup}

\indent All the models discussed in this work are two-dimensional in space with cylindrical symmetry, and are defined by a disk radius $R_{\textrm{max}}$ and a halo half-height $z_{\textrm{max}}$, with a cell size of 50\,pc in both dimensions. The particle energy grid runs from 100\mev\ to 1\pev\ with ten bins per decade. The transport of CRs arises from diffusion for the spatial part, and includes ionisation, synchrotron, Bremsstrahlung, inverse-Compton scattering, and hadronic interaction losses for the momentum part. For the latter, the production of secondary electrons and positrons is taken into account and included in the complete transport and radiation calculation. All solutions correspond to a steady state. For more details about the implementation of the physical processes and the inner workings of the code, see the GALPROP Explanatory Supplement available at the GALPROP website.

\indent Four different sizes were considered for the synthetic star-forming disk galaxies: ($R_{\textrm{max}}$, $z_{\textrm{max}}$)=(2\,kpc, 1\,kpc), (5\,kpc, 1\,kpc), (10\,kpc, 2\,kpc), and (20\,kpc, 4\,kpc). The larger size corresponds to the dimensions of the Milky Way in the reference GALPROP run, and I kept the same aspect ratio for the smaller sizes, except for the smaller model where it would have resulted in a too small halo. The size of the halo for the Milky Way and how it evolves with galaxy size are unsettled questions, therefore this assumption of a nearly constant aspect ratio in galaxy dimensions is challenged below (see Sects. \ref{res_params}, \ref{res_fermilat}, and \ref{res_radio}).

\indent For each size, the galaxy consisted of a disk of atomic gas with an exponential distribution in $R$ and $z$, with scale lengths of $R_{\textrm{H}}=R_{\textrm{max}}/2$ and $z_{\textrm{H}}=$100\,pc respectively, and a peak gas density of $n_{\textrm{H}}=$2\,H\dunit. On top of that, different molecular gas distributions were implemented: cores of radius $R_{\textrm{core}}=$100, 200, 500\,pc, and 1\,kpc, and two rings with inner and outer radius $(R_{i}, R_{o})=(0.2 \, R_{\textrm{max}},0.3 \, R_{\textrm{max}})$ and $(0.4 \, R_{\textrm{max}},0.6 \, R_{\textrm{max}})$, all distributions extending 100\,pc on either side of the galactic plane (it was verified that using a smaller molecular gas scale height of 50\,pc does not alter the trends in the evolution of the \gray\ luminosity with galactic properties). Not all combinations of molecular gas distributions and galaxy sizes were tested, see the list of models summarised in Table \ref{tab_modellist}. Then, for each molecular gas profile, seven gas densities were tested: $n_{\textrm{H}_2}=$5, 10, 20, 50, 100, 200, and 500\,H$_2$\dunit. The initial sample of models includes $14 \times 7=88$ synthetic star-forming galaxies, not counting variations of some parameters to assess their importance. While some configurations may be considered as good models for existing systems such as the Milky Way or M82, others may be less realistic, such as the ($R_{\textrm{max}}$, $z_{\textrm{max}}$)=(20\,kpc, 4\,kpc) model with a large ring of $n_{\textrm{H}_2}=$500\,H$_2$\dunit gas, yielding an SFR of $\sim10^4$\wunit. These were included to illustrate the effects of changing galactic parameters. 

\subsection{Time scales}
\label{model_timescales}

\indent To facilitate the interpretation of the results presented in the following, I provide the typical time scales for the various processes involved in the transport of CRs. These hold for the assumptions used in the star-forming galaxy model (such as the spatial diffusion coefficient).
\begin{align}
\tau_{\textrm{diff}} &= 2.8\,10^{14} \, \left( \frac{E}{4\,\textrm{GeV}} \right)^{-0.5} \, \left( \frac{z}{1\,\textrm{kpc}} \right)^2 \, \textrm{s} \\
\tau_{\textrm{adv}} &= 3.1\,10^{14} \, \left( \frac{z}{1\,\textrm{kpc}} \right) \, \left( \frac{V}{100\,\textrm{km}\,\textrm{s}^{-1}} \right)^{-1} \, \textrm{s} \\
\tau_{\textrm{pp}} &= 2.2\,10^{15} \, \left( \frac{n_{\textrm{H}}}{1\,\textrm{cm}^{-3}} \right)^{-1} \, \textrm{s} \\
\tau_{\textrm{Br}} &= 8.9\,10^{14} \, \left( \frac{n_{\textrm{H}}}{1\,\textrm{cm}^{-3}} \right)^{-1} \, \textrm{s} \\
\tau_{\textrm{IC}} &= 9.9\,10^{15} \, \left( \frac{E}{1\,\textrm{GeV}} \right)^{-1} \, \left( \frac{U_{\textrm{rad}}}{1\,\textrm{eV}\,\textrm{cm}^{-3}} \right)^{-1} \, \textrm{s}  \\
\tau_{\textrm{Synch}} &= 9.9\,10^{15} \, \left( \frac{E}{1\,\textrm{GeV}} \right)^{-1} \, \left( \frac{U_{\textrm{mag}}}{1\,\textrm{eV}\,\textrm{cm}^{-3}} \right)^{-1} \, \textrm{s} \\
\tau_{\textrm{ion}} &= 4.8\,10^{15} \, \left( \frac{E}{1\,\textrm{GeV}} \right) \, \left( \frac{n_{\textrm{H}}}{1\,\textrm{cm}^{-3}} \right)^{-1} \, \textrm{s}
\end{align}
The time scales are those of the diffusion, advection, hadronic interaction, Bremsstrahlung, inverse-Compton, synchrotron, and ionisation processes. For the spatial transport processes, the typical length was taken at 1\,kpc, which is the order of magnitude for the galactic halo size. The dependence on energy for the diffusion time scale applies only for $E > 4$\gev. The hadronic interaction time scale was computed for an energy of 1\gev, with a cross-section of 30\,mbarn and an inelasticity parameter of 0.5 \citep[note that the cross-section increases slightly to reach 40\,mbarn at 1\tev, see][]{Kelner:2006}. The Bremsstrahlung-loss time scale corresponds to an electron energy of 1\gev, and typically has a logarithmic dependence on energy that was neglected in the above formula \citep{Schlickeiser:2002}. The inverse-Compton-loss time scale was computed in the Thomson approximation, with $U_{\textrm{rad}}$ corresponding to the radiation energy density. The synchrotron-loss time scale is the same formula, with $U_{\textrm{mag}}$ corresponding to the magnetic energy density. The ionisation-loss time scale was computed for protons in the ISM, using the lowest value of the loss rate (which is obtained for a Lorentz factor of $\sim3$) and neglecting its energy dependence (which leads to variation by a factor of at most 3 over the 100\mev-10\gev\ range).

\indent Because the ISRF and magnetic field components are deduced from the gas distribution in my model, it is useful to recast some of the above formulae as a function of molecular gas density. For a given gas distribution, such as a molecular core or a ring of a given extent on top of the atomic disk, the molecular gas density is the independent variable that drives the non-thermal properties. Accordingly, from Eqs. \ref{eq_bfield} and \ref{eq_ir}, and using a thickness of 200\,pc for the molecular gas, one obtains
\begin{align}
\tau_{\textrm{pp}} &= 1.1\,10^{15} \, \left( \frac{n_{\textrm{H}_2}}{1\,\textrm{cm}^{-3}} \right)^{-1} \, \textrm{s} \\
\tau_{\textrm{Br}} &= 4.5\,10^{14} \, \left( \frac{n_{\textrm{H}_2}}{1\,\textrm{cm}^{-3}} \right)^{-1} \, \textrm{s} \\
\tau_{\textrm{IC}} &= 1.8\,10^{16} \, \left( \frac{E}{1\,\textrm{GeV}} \right)^{-1} \, \left( \frac{n_{\textrm{H}_2}}{1\,\textrm{cm}^{-3}} \right)^{-1.4} \, \textrm{s}  \\
\tau_{\textrm{Synch}} &= 5.0\,10^{15} \, \left( \frac{E}{1\,\textrm{GeV}} \right)^{-1} \, \left( \frac{n_{\textrm{H}_2}}{1\,\textrm{cm}^{-3}} \right)^{-1.2} \, \textrm{s} \\
\tau_{\textrm{ion}} &= 2.4\,10^{15} \, \left( \frac{E}{1\,\textrm{GeV}} \right) \, \left( \frac{n_{\textrm{H}_2}}{1\,\textrm{cm}^{-3}} \right)^{-1} \, \textrm{s} ,
\end{align}
where only the infrared component of the ISRF was considered in the inverse-Compton time scale (it is by far the dominant component above $n_{\textrm{H}_2}=$10-20\,H$_2$\dunit).

\section{Results}
\label{res}

\indent In this section, I present the results for the synthetic star-forming galaxies introduced above. To allow the connection with observations, the \gray\ luminosities are often given in three spectral bands: 100\kev-10\mev\ (band I, soft gamma-rays), 100\mev-100\gev\ (band II, high-energy gamma-rays), 100\gev-100\tev\ (band III, very-high-energy gamma-rays). Examples of current instruments covering these bands are INTEGRAL/SPI, Fermi/LAT, and HESS, respectively.

\indent I first focus on a subset of the results to illustrate how gas distribution and density affect the \gray\ emission, and how the latter scales with global parameters. This is followed by a discussion of the impact of changing some parameters of the model. I then discuss the complete set of results in the context of the population study performed from \textit{Fermi}/LAT observations, and of the far-infrared - radio correlation.

\subsection{Gamma-ray luminosities and spectra}
\label{res_gamma}

\begin{figure}[H]
\begin{center}
\includegraphics[width= \columnwidth]{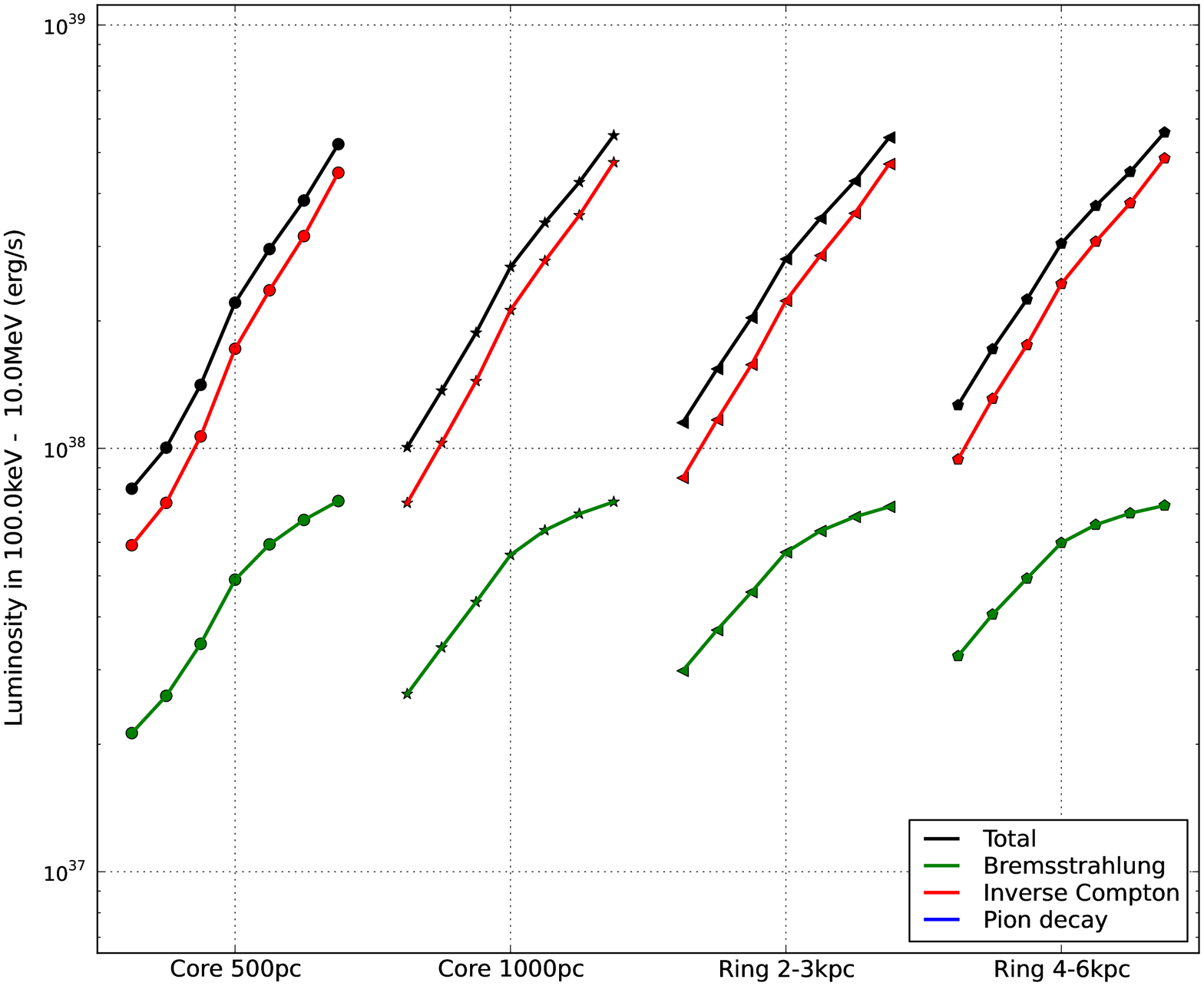}
\includegraphics[width= \columnwidth]{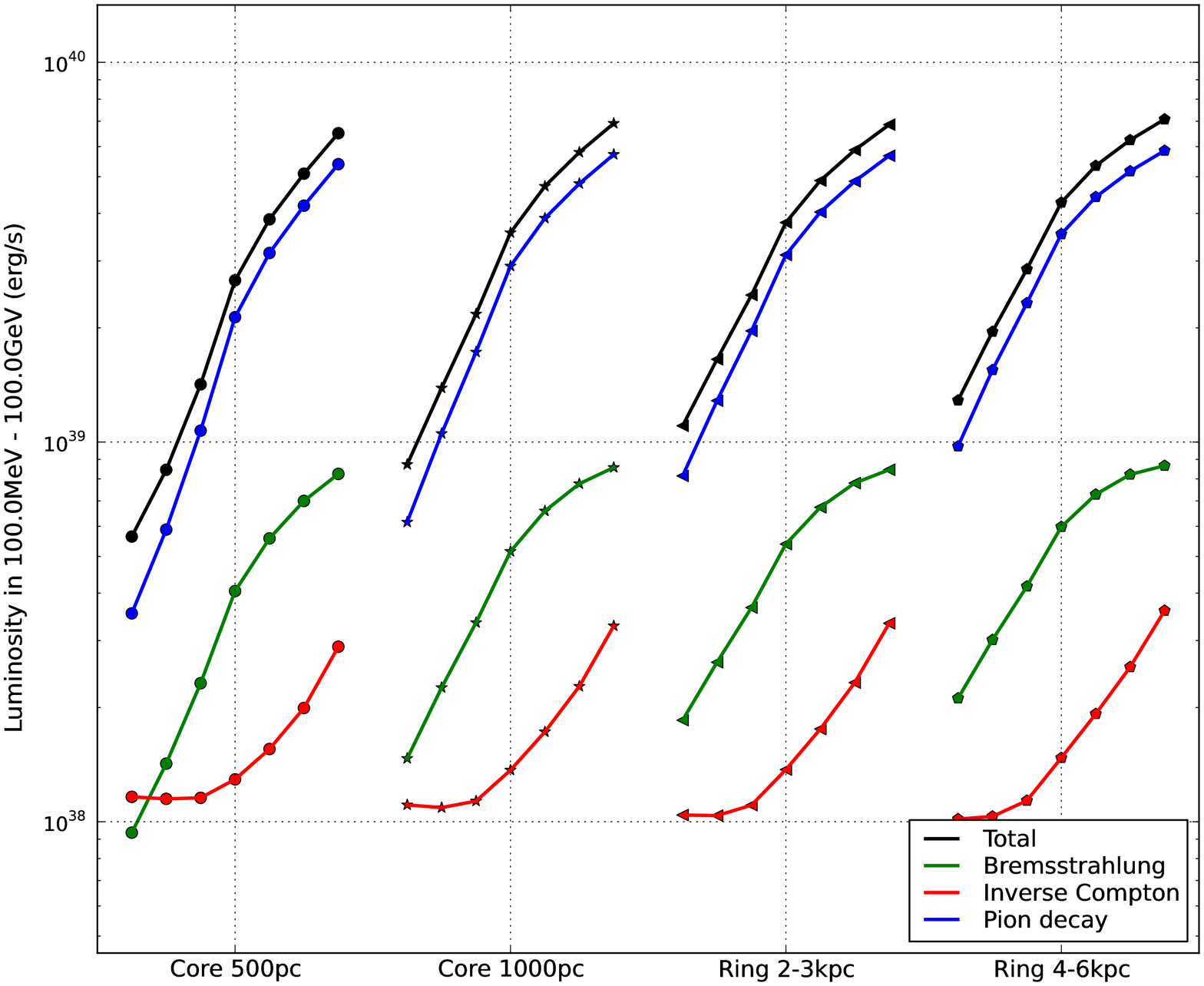}
\includegraphics[width= \columnwidth]{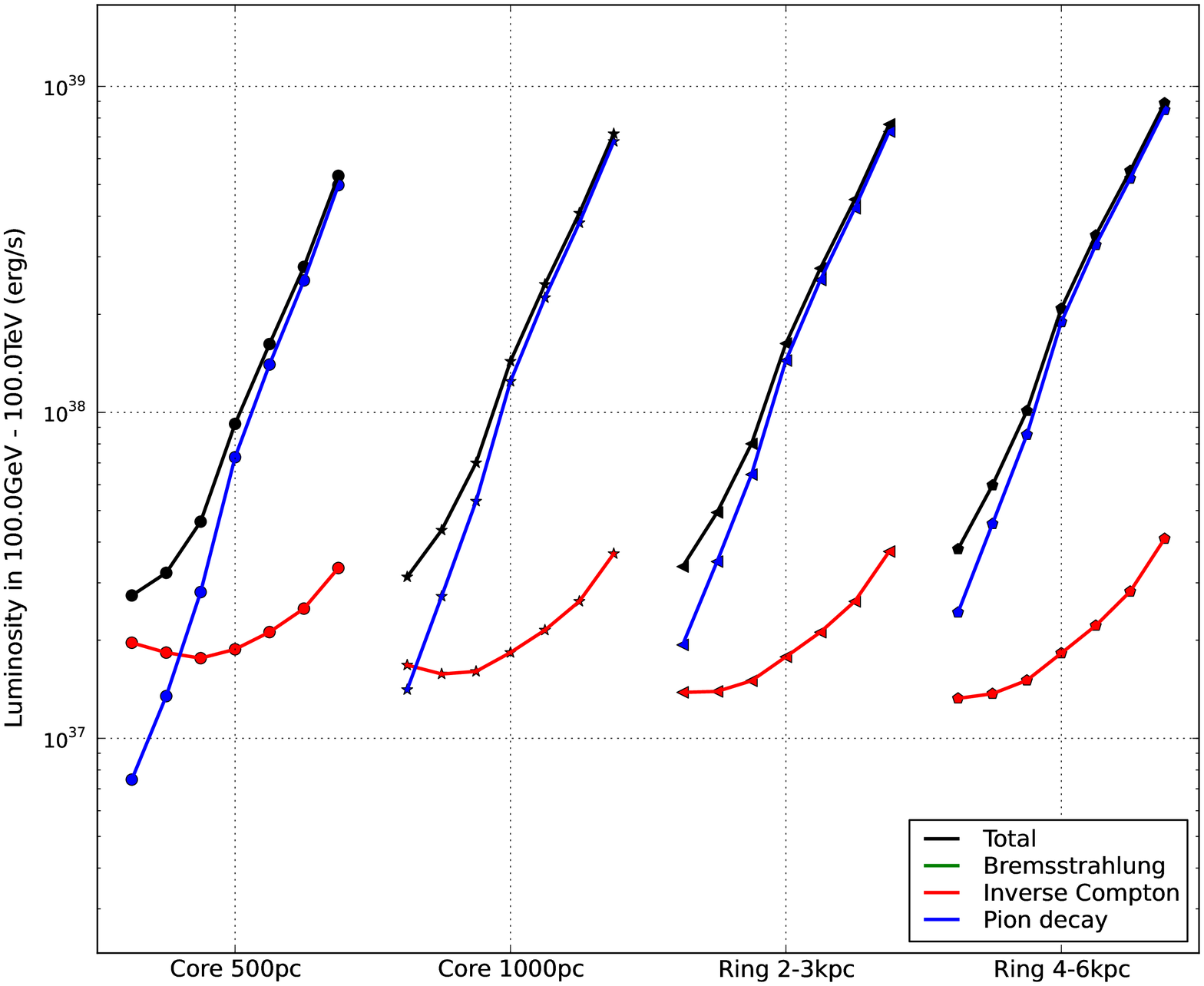}
\caption{Luminosity in 3 \gray\ bands for a star-forming disk galaxy with ($R_{\textrm{max}}$, $z_{\textrm{max}}$)=(10\,kpc, 2\,kpc), 4 profiles of molecular gas (given in abscissa), and 7 molecular gas densities for each profile ($n_{\textrm{H}_2}=$5, 10, 20, 50, 100, 200, 500\,H$_2$\dunit). To illustrate the effects of the gas distribution, the \gray\ luminosities are given for the same input cosmic-ray luminosity.}
\label{fig_lumgamma_bands_noscale}
\end{center}
\end{figure}

\begin{figure}[H]
\begin{center}
\includegraphics[width= \columnwidth]{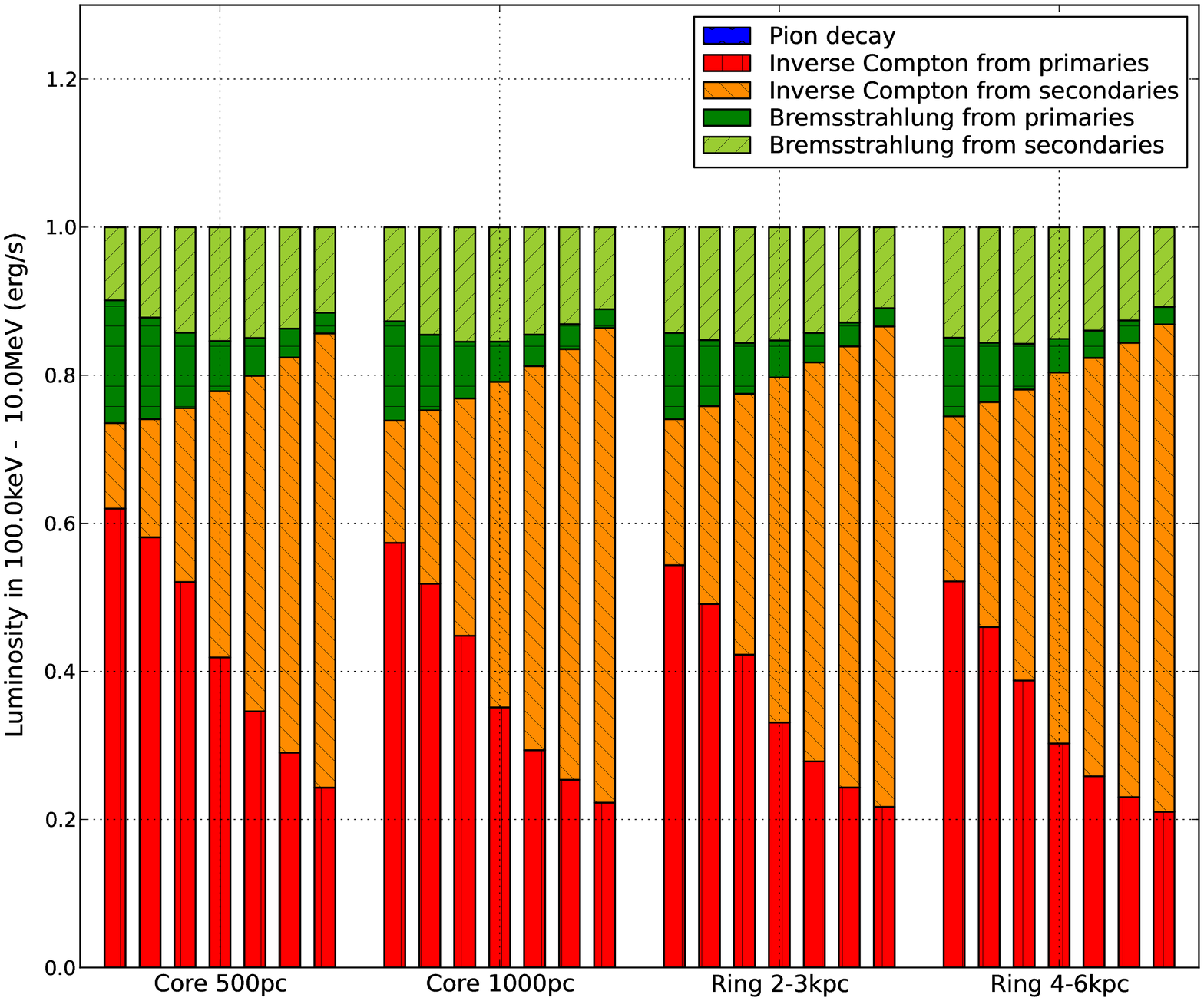}
\includegraphics[width= \columnwidth]{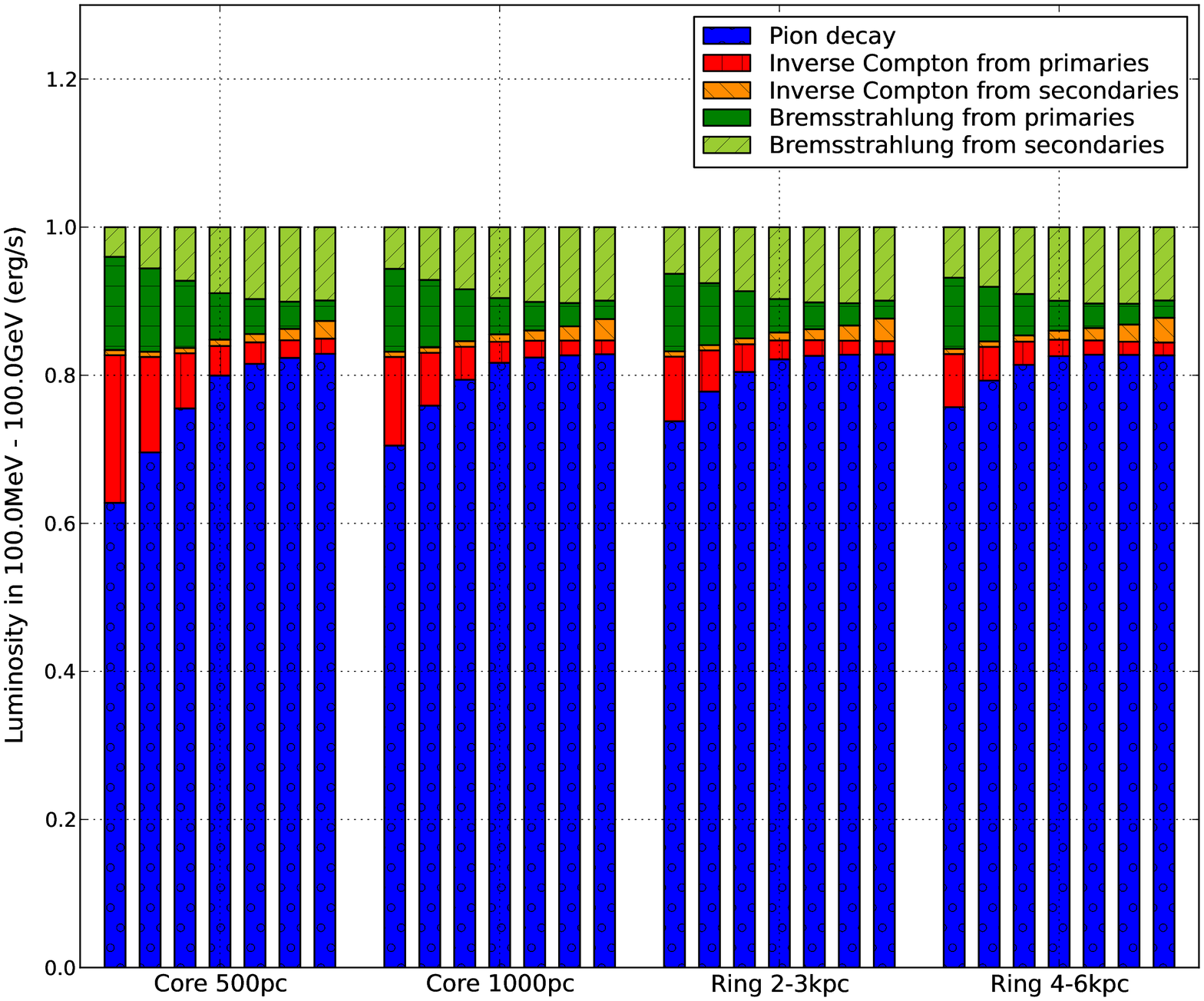}
\includegraphics[width= \columnwidth]{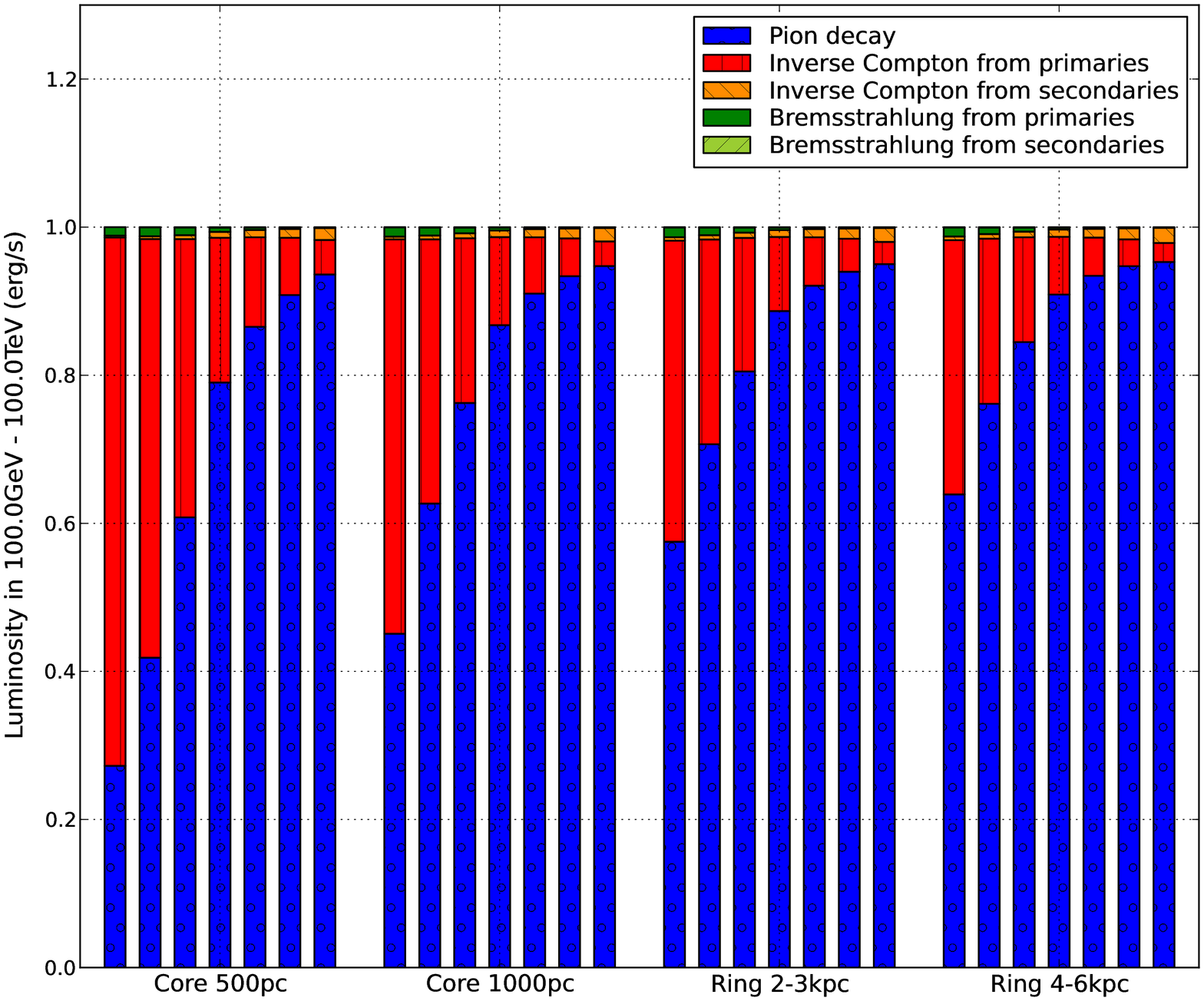}
\caption{Distribution of the luminosity in 3 \gray\ bands, in terms of physical processes and emitting particles, for a star-forming disk galaxy with ($R_{\textrm{max}}$, $z_{\textrm{max}}$)=(10\,kpc, 2\,kpc), 4 profiles of molecular gas (given in abscissa), and 7 molecular gas densities for each profile ($n_{\textrm{H}_2}=$5, 10, 20, 50, 100, 200, 500\,H$_2$\dunit).}
\label{fig_distribgamma_bands_noscale}
\end{center}
\end{figure}

\indent To illustrate the effects of the gas distribution and density on \gray\ emission in a star-forming galaxy, I restrict the discussion to the case of a disk galaxy with $R_{\textrm{max}}=$10\,kpc for the four molecular gas profiles and seven molecular gas densities introduced above. To allow a clear comparison, the runs were performed for the same input CR luminosity (which is that of the Milky Way, see Sect. \ref{model_src}). The plots of Fig. \ref{fig_lumgamma_bands_noscale} show the evolution of the \gray\ luminosity in three different bands for this particular setup, while the plots of Fig. \ref{fig_distribgamma_bands_noscale} show the distribution of the emission in terms of physical processes and emitting particles (primary electrons versus secondary electrons and positrons, hereafter primaries and secondaries for short).

\indent These plots do not only show the effect of increasing gas density, and correspondingly of the enhanced energy losses, but also include an effect of the distribution of CR sources. Indeed, from the smaller core to the larger ring, and with increasing density, the ratio of star formation occurring in dense molecular regions is higher than that distributed in the lower-density atomic disk. This is why for a same increase in molecular gas density, the 500\,pc core configuration exhibits the highest contrasts between the low-density case ($n_{\textrm{H}_2}=5$\,H$_2$\dunit) and the high-density case ($n_{\textrm{H}_2}=500$\,H$_2$\dunit): the luminosities increase because of a higher conversion efficiency of particle energy to radiation, which arises from a higher gas density and from a larger proportion of CRs being injected in a higher-density medium. 

\indent The evolution of the luminosity in all three \gray\ bands, for an increase of a factor 100 of the molecular gas density, shows the following trends (I recall that this discussion holds for the same input CR luminosity):

\indent Band I (100\kev-10\mev): The emission is always dominated by IC, with Bremsstrahlung contributing at the 15-25\% level depending on gas density. The IC from primaries is almost constant, which is consistent with primaries being IC loss-dominated (the increase in the radiation field density is balanced by a decrease in the steady-state primaries population). The IC from secondaries increases by a factor of about 10, which is due to a rise of the production rate of secondaries (the production rate increases not as much as the gas density because low-energy CRp progressively move from a diffusion-dominated regime to a loss-dominated regime). The net result is that emission in band I gradually becomes dominated by secondaries, at a level of 70-80\% for the highest gas density \citep[for observations and interpretation of the diffuse soft \gray\ emission from the Milky Way, see][]{Porter:2008,Bouchet:2011}. Overall, this contribution of secondaries increases from the very-high-energy band to the soft-gamma-ray band (band III to I), which reflects mainly that secondaries are the particles with the steepest spectrum. 

\indent Band II (100\mev-100\gev): The emission is always dominated by pion decay, while leptonic processes contribute at the 20-30\% level, with an increasing fraction of it coming from secondaries (eventually reaching 75\% for the highest gas densities). The evolution of the pion decay and Bremsstrahlung emission with gas density flattens, and this is due to a transition in the transport regime. At the lowest gas densities, diffusion is an important process at intermediate particle energies $\sim$1-10\gev, so the steady-state particle population is not strongly affected by gas density and the luminosity of gas-related components increases with gas density, although slower than linearly. At the highest gas densities, Bremsstrahlung losses (for CR leptons) and hadronic losses (for CR nuclei) seriously compete with diffusion, therefore the luminosity of gas-related components rises more weakly with gas density as the transport is moving towards a loss-dominated regime. Note that the shape of this flattening probably depends on the diffusion conditions (a higher diffusion coefficient at low energies would imply a flattening occurring later, at a higher density). 

\indent Band III (100\gev-100\tev): The emission is dominated by pion decay, except for the lowest average gas densities where it is either dominated by IC or where IC is within a factor of 2-3. Inverse-Compton emission is here always dominated by primaries, and this contribution is constant with gas density because the steady-state primary CRe population is loss-dominated. In contrast, the contribution from secondaries will rise with gas density, following the increase of the production rate of secondaries from high-energy CRp. But because secondaries have a softer injection spectrum, and because the parent CRp population gradually suffers from losses, secondaries never dominate primaries at these high energies, and the increase of the total IC emission remains moderate, only by a factor of 2-3. In contrast, pion decay emission increases faster with gas density. 
\newpage
\newpage
\begin{figure}[H]
\begin{center}
\includegraphics[width= 8.25cm]{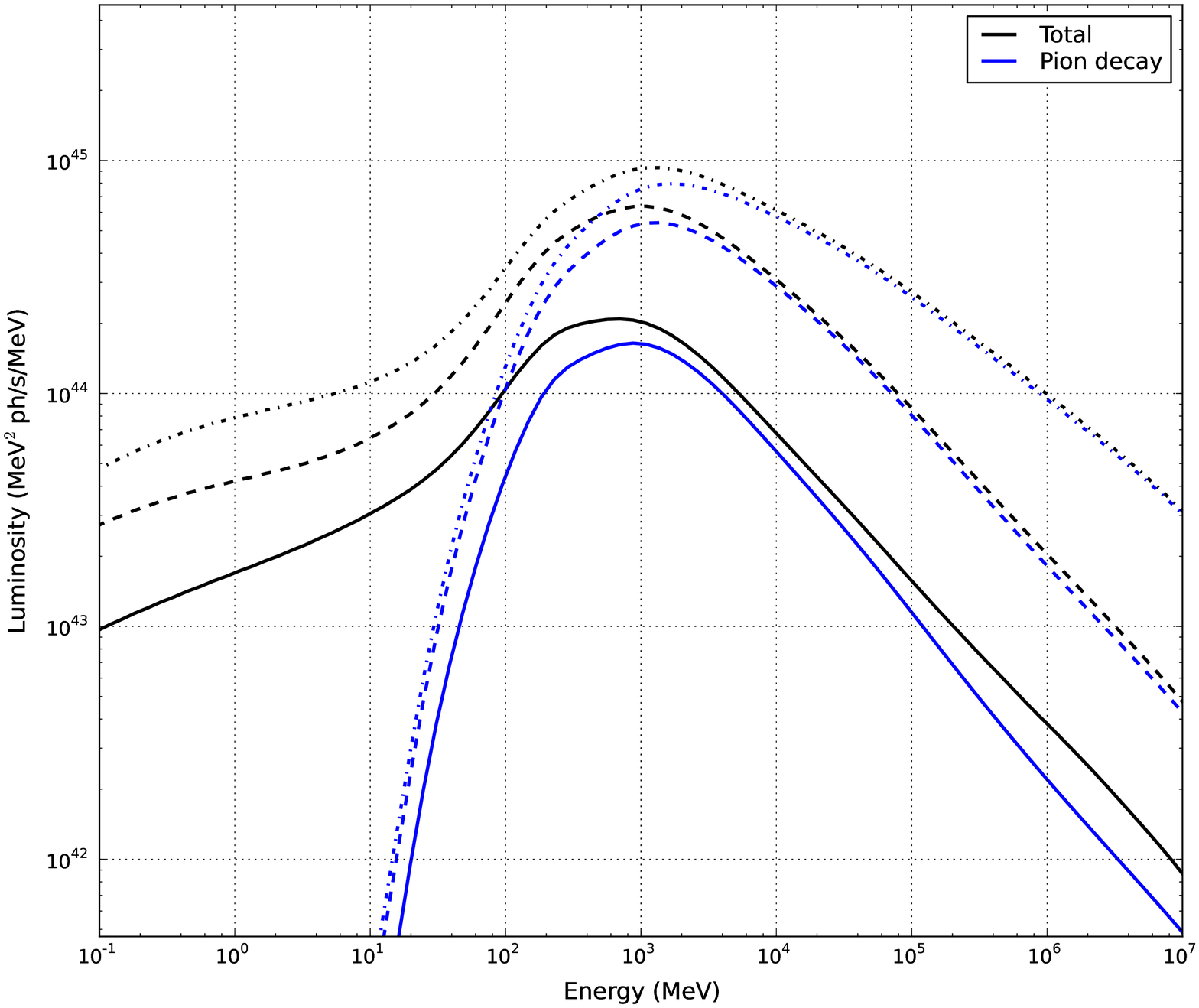}
\includegraphics[width= 8.25cm]{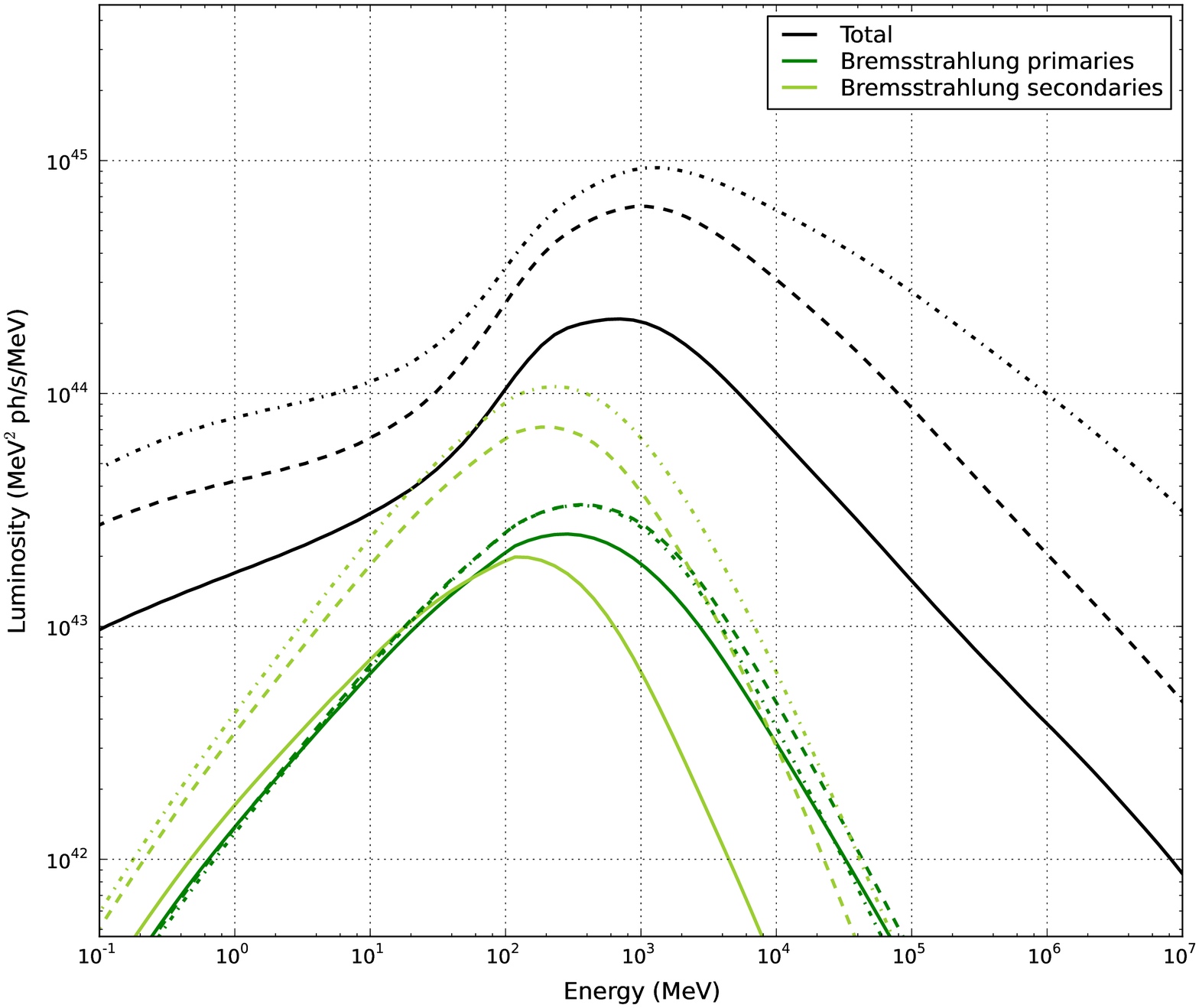}
\includegraphics[width= 8.25cm]{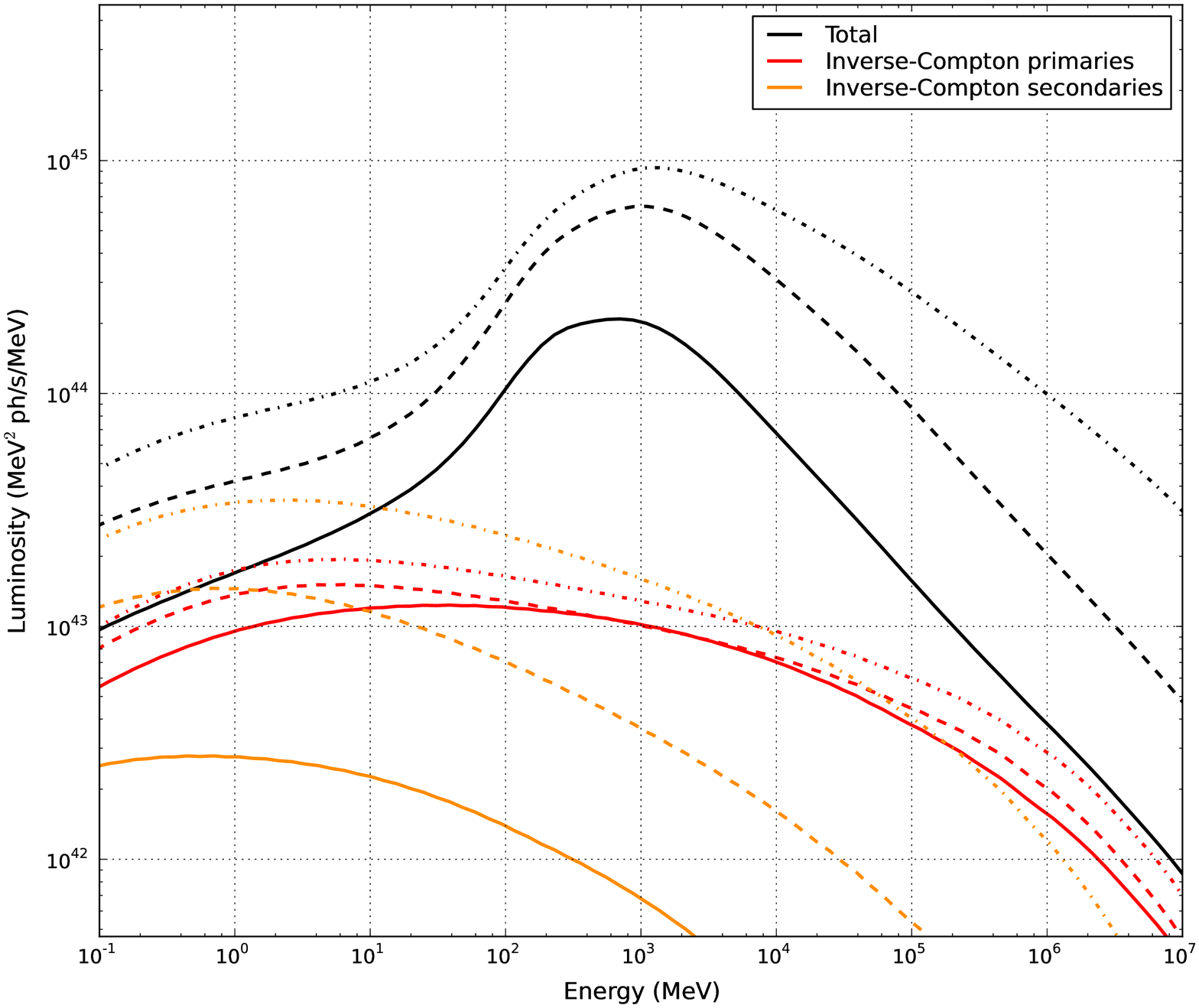}
\caption{Spectra of pion decay (top), Bremsstrahlung (middle), and inverse-Compton (bottom) emission for a star-forming disk galaxy with ($R_{\textrm{max}}$, $z_{\textrm{max}}$)=(10\,kpc, 2\,kpc) and a 4-6\,kpc ring distribution of molecular gas. The solid, dashed, and dot-dashed curves correspond to $n_{\textrm{H}_2}$=5, 50, and 500\,H$_2$\dunit, respectively. For the leptonic processes, the contribution from primary electrons and secondary positrons are shown. In all plots, the total \gray\ emission from all processes and all particles is shown in black.}
\label{fig_specgamma}
\end{center}
\end{figure}
\begin{figure}[H]
\begin{center}
\includegraphics[width= 8.25cm]{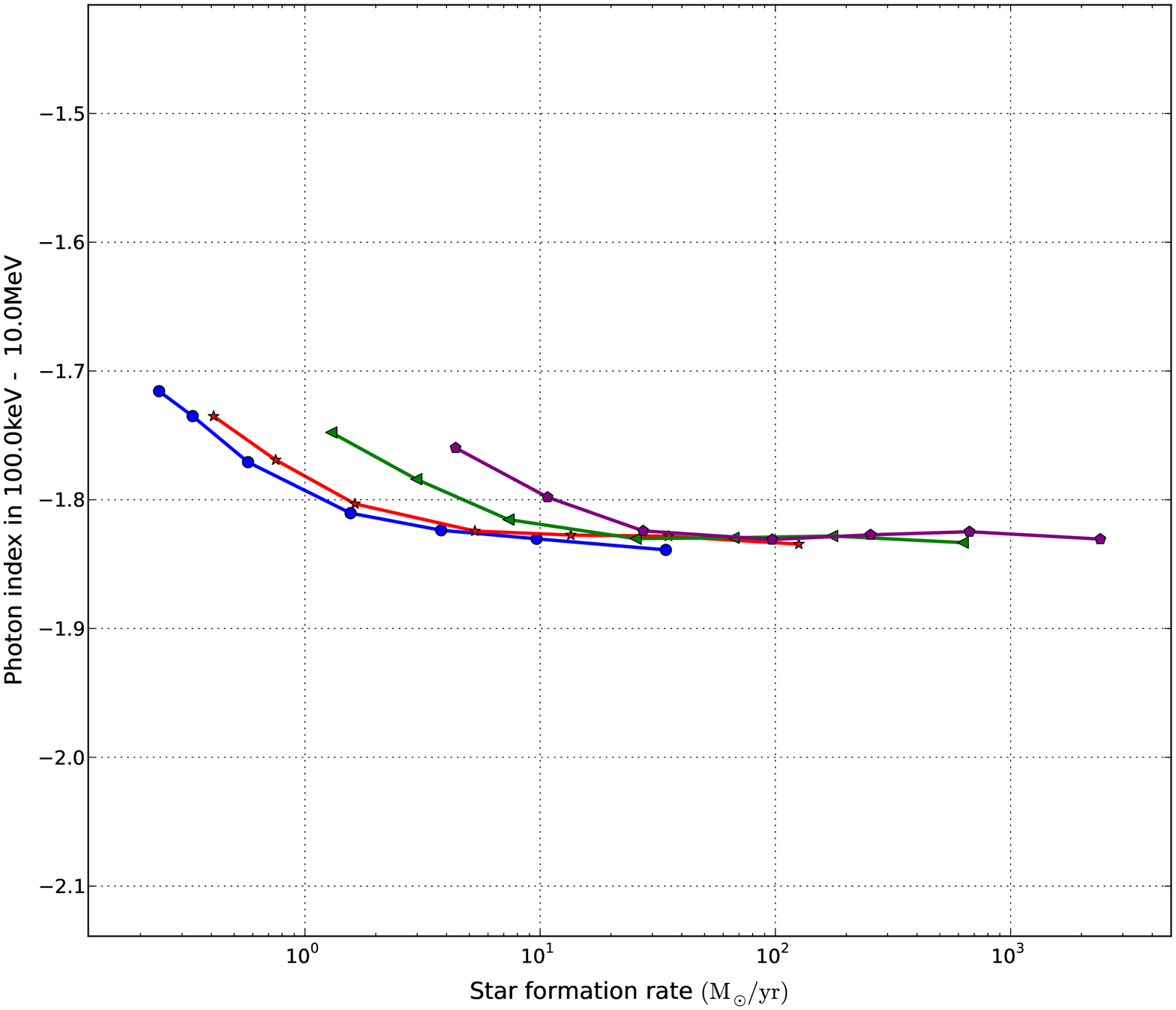}
\includegraphics[width= 8.25cm]{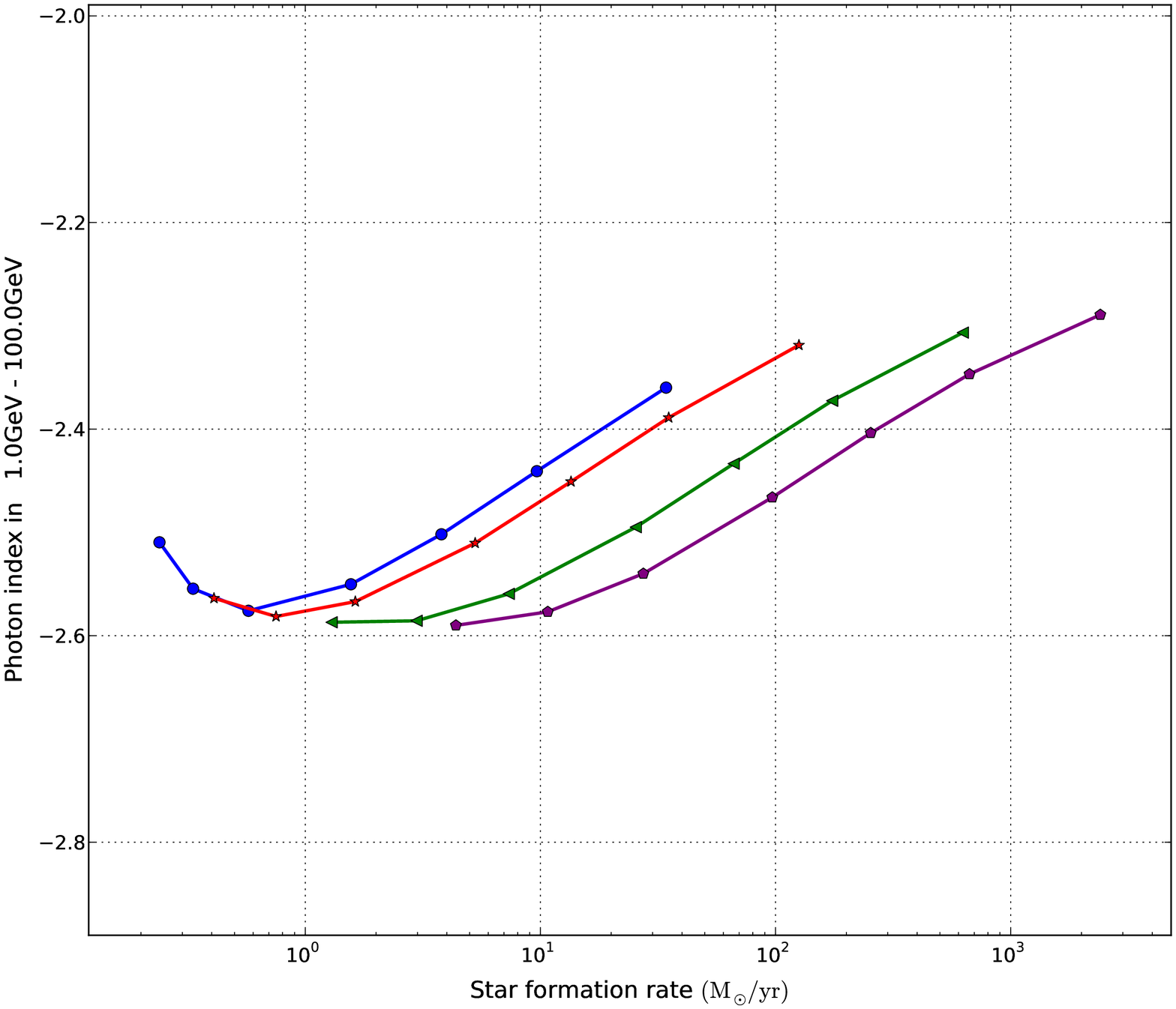}
\includegraphics[width= 8.25cm]{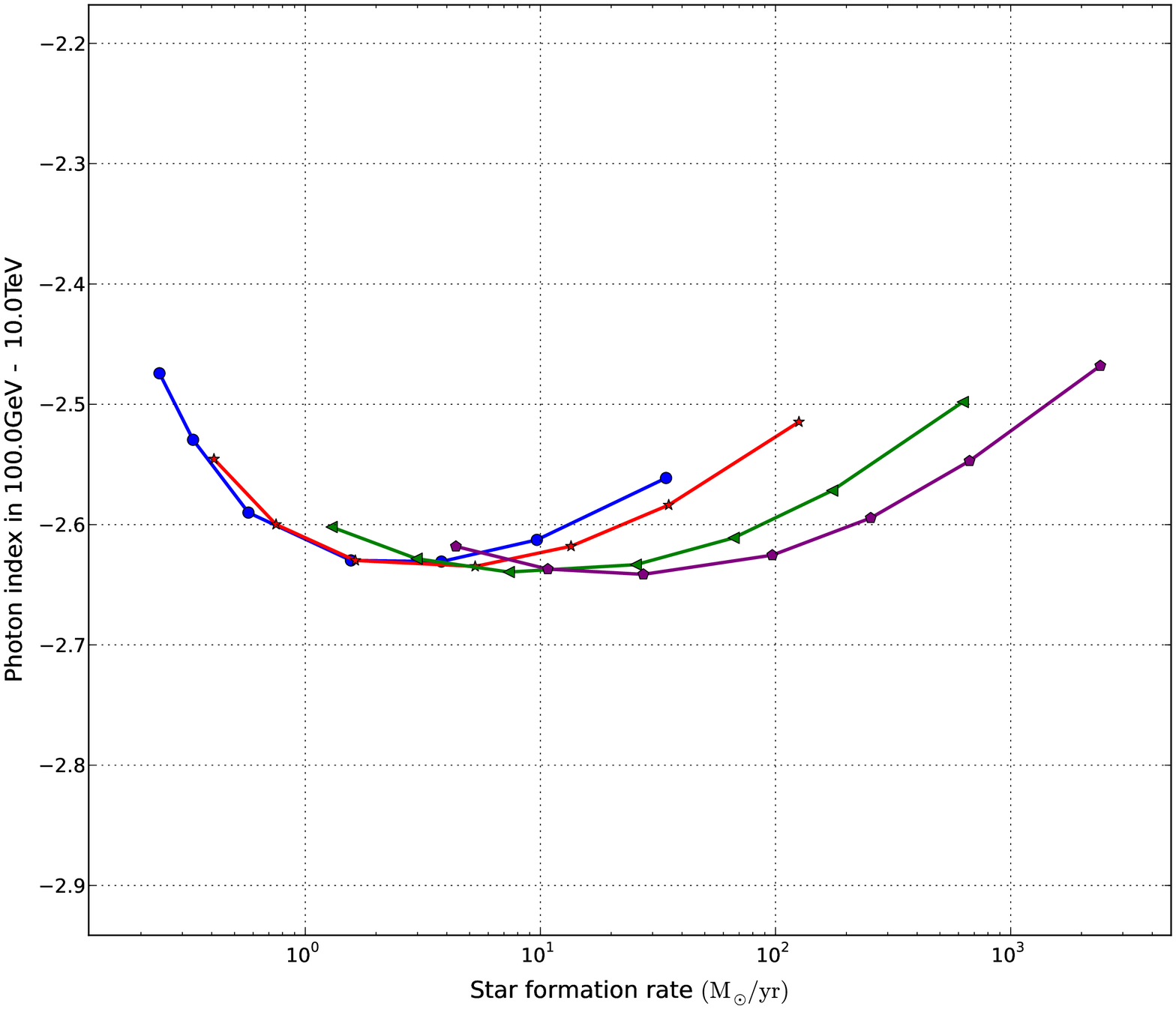}
\caption{Photon spectral indices over 3 \gray\ bands for a star-forming disk galaxy with ($R_{\textrm{max}}$, $z_{\textrm{max}}$)=(10\,kpc, 2\,kpc), 4 profiles of molecular gas, and 7 molecular gas densities for each profile ($n_{\textrm{H}_2}=$5, 10, 20, 50, 100, 200, 500\,H$_2$\dunit). Blue, red, green, and purple curves correspond to the 500\,pc core, 1\,kpc core, 2-3\,kpc ring, and 4-6\,kpc ring gas profiles, respectively. Note that the energy bands differ from those used for the luminosities.}
\label{fig_idxgamma}
\end{center}
\end{figure}
\newpage
For the lowest gas densities, the increase is almost linear because high-energy CRp emitting in this band are close to diffusion-dominated. This trend flattens with further increase of the gas density, however, because pion decay losses become strong enough to reduce the steady-state CRp population. The effect on the pion decay emission extends well into the 100\gev-1\tev\ range.

\indent The plots of Fig. \ref{fig_specgamma} show the \gray\ spectra of each emission process independently, for the large ring distribution with gas densities $n_{\textrm{H}_2} =$5, 50, and 500\,cm${-3}$, and still for the same input CR power. The respective contribution of primaries and secondaries is given. These plots illustrate some of the statements made above, in the discussion of the luminosity evolution in the three \gray\ bands. One can see the flattening of the pion decay emission in the 100\mev-1\tev\ band as gas density increases, the relative steadiness of the primary leptonic components, and the growth of the secondary leptonic contribution.

\indent The plots of Fig. \ref{fig_idxgamma} show the photon spectral indices over three spectral bands that differ from those used before for the luminosities (choices made so that a power-law approximation can hold). These are given for the disk galaxy model with $R_{\textrm{max}}=$10\,kpc, for the four molecular gas profiles and seven molecular gas densities, as a function of SFR. The flattening of the spectrum with increasing gas density is clearly apparent for band II. The emission in band III exhibits that same feature but also flattens at low SFRs due to the predominance of inverse-Compton emission.

\begin{figure}[!ht]
\begin{center}
\includegraphics[width= 8.25cm]{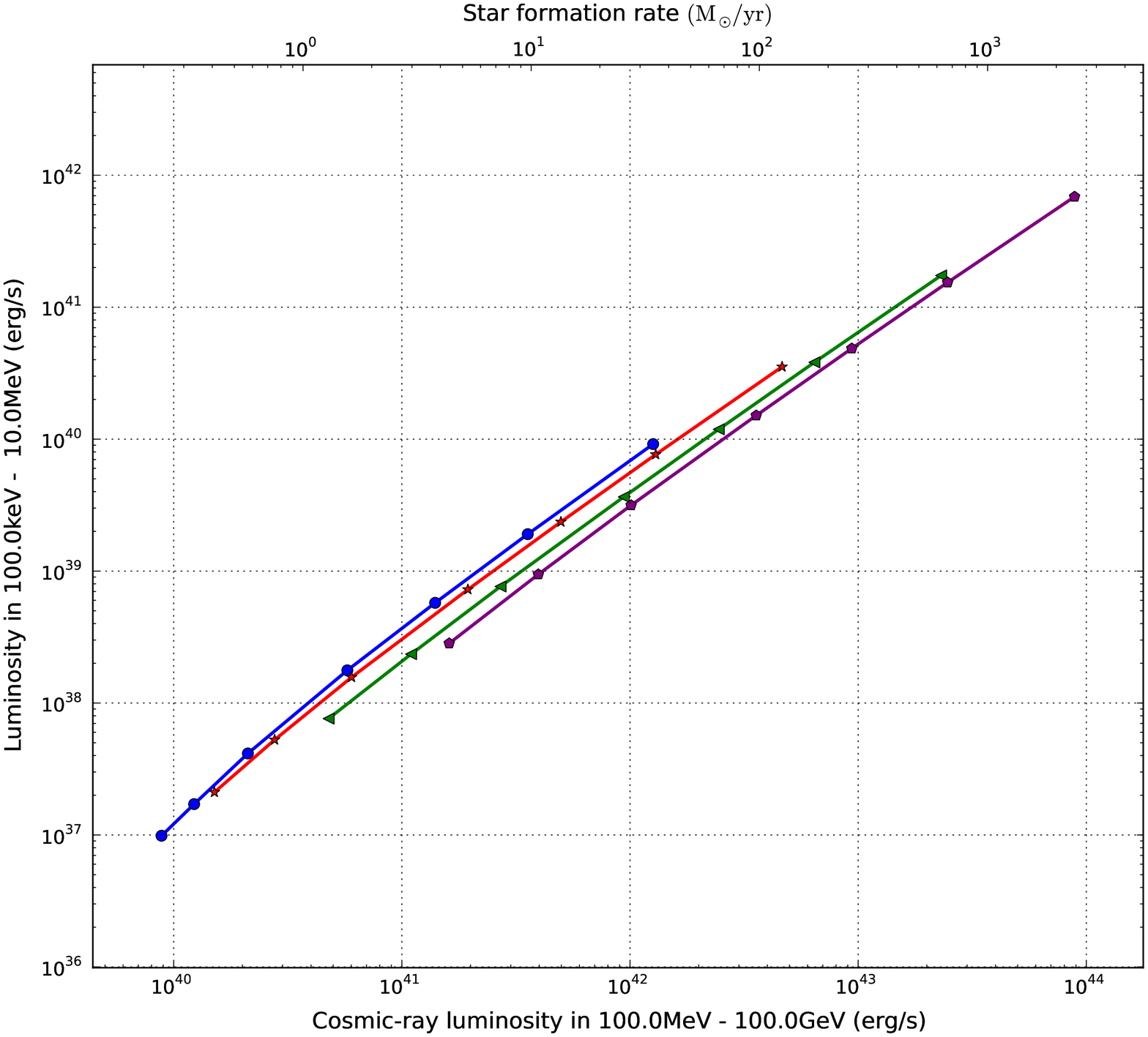}
\includegraphics[width= 8.25cm]{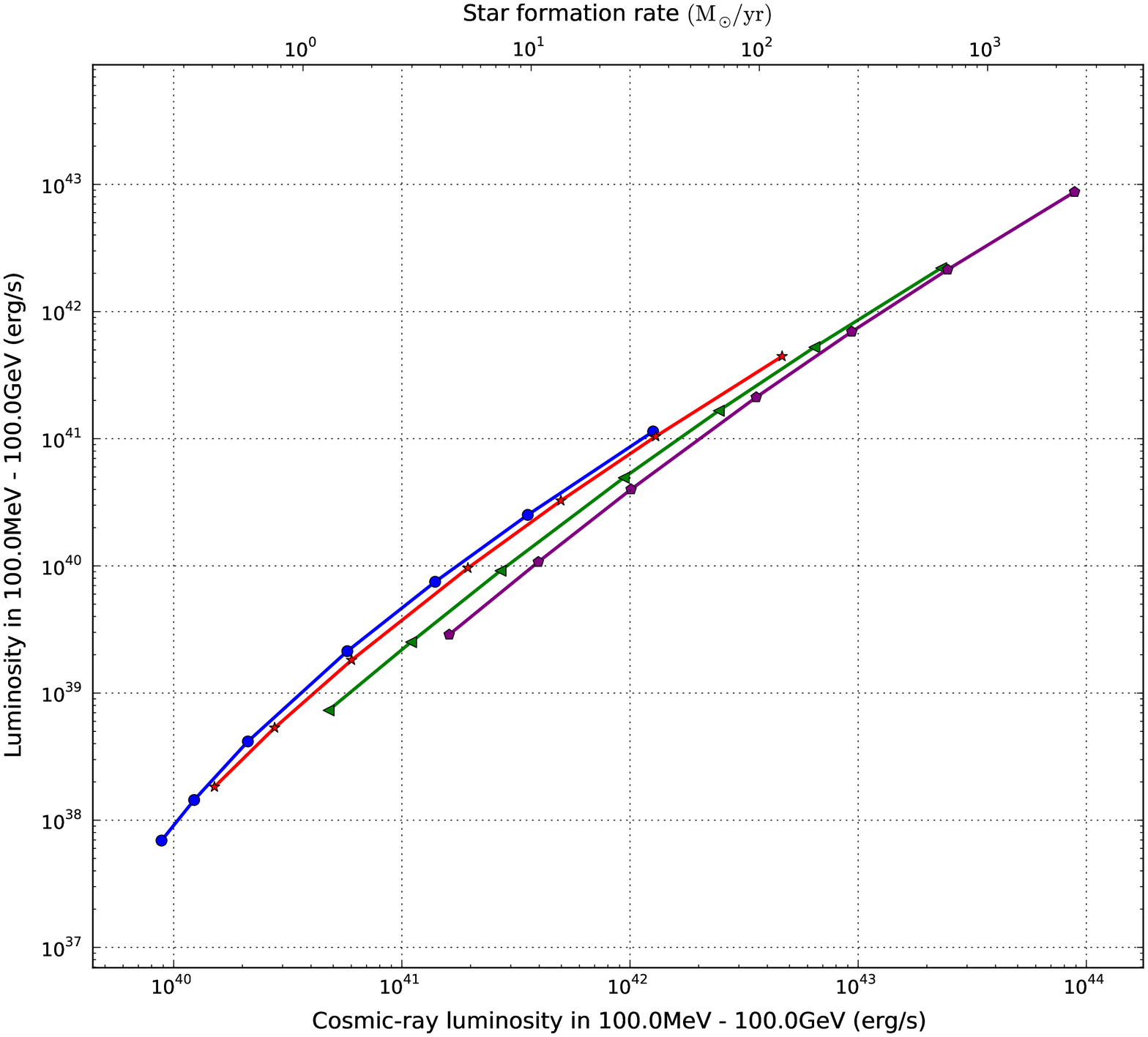}
\includegraphics[width= 8.25cm]{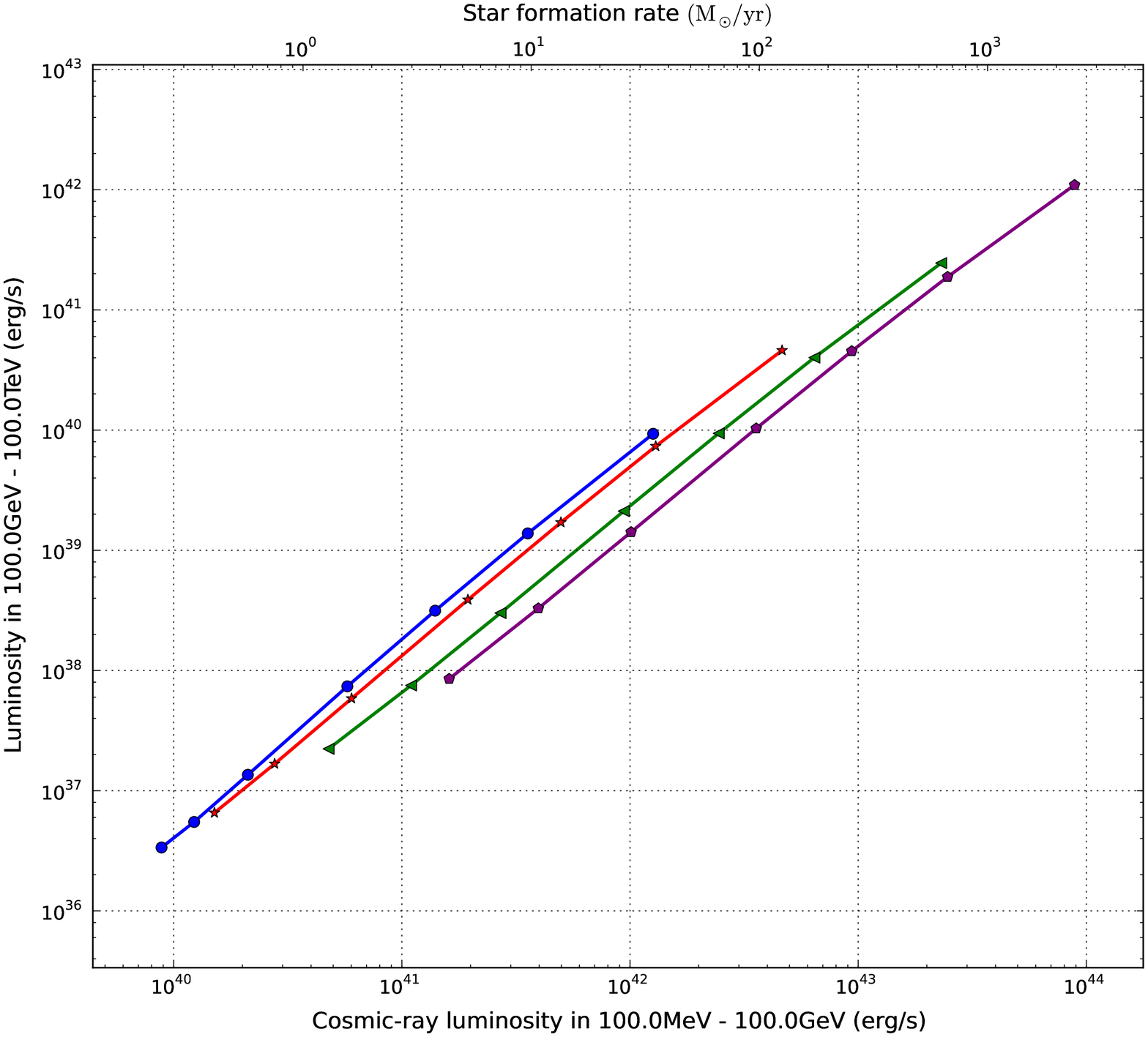}
\caption{Luminosity in 3 \gray\ bands  as a function of star formation rate and cosmic-ray power, for a star-forming disk galaxy with ($R_{\textrm{max}}$, $z_{\textrm{max}}$)=(10\,kpc, 2\,kpc), 4 profiles of molecular gas (given in abscissa), and 7 molecular gas densities for each profile ($n_{\textrm{H}_2}=$5, 10, 20, 50, 100, 200, 500\,H$_2$\dunit). The cosmic-ray input power generated by the galaxy is assumed to scale with total star formation rate, using the Milky Way as normalisation.}
\label{fig_lumgamma_bands_scalesfr}
\end{center}
\end{figure}

\subsection{Scaling of luminosities with global properties}
\label{res_scaling}

\indent Under the assumption that the CR power generated by a star-forming galaxy is only proportional to its total SFR, and that the latter can be described from the Schmidt-Kennicutt relation, total luminosities were computed for all models in the sample. In the following, I present the theoretical scaling of the luminosity in the three \gray\ spectral bands in a qualitative and quantitative way. For clarity, the plots only illustrate the results for the disk galaxy with $R_{\textrm{max}}=$10\,kpc for the four molecular gas profiles and seven molecular gas densities introduced previously. The complete sample is shown for band II in a following subsection, because it is the only band where a population study was recently performed.

\indent The plots of Fig. \ref{fig_lumgamma_bands_scalesfr} show the evolution of \gray\ luminosity in the three spectral bands as a function of SFR and CR power. For each gas distribution (the two cores and two rings), the parameters of a scaling law were computed from the high gas density cases, under the hypothesis of a scaling of the form
\begin{equation}
\log L_{\gamma} =  \beta \log P_{\textrm{CR}} + \delta 
\end{equation}
Most of the discussion below focuses on the index $\beta$ because this contains most of the physics of the transport, while the offset $\delta$ is more related to normalisation considerations.

\indent Band I (100\kev-10\mev): The luminosity at high gas densities evolves with an index $\delta \sim$1.2-1.3. At lower gas densities, the luminosity decreases faster than that because an increasing fraction of the CR power is released into the low-density atomic disk (the effect is stronger for the small-core model). The more-than-linear increase is ultimately due to the contribution of secondaries to the IC emission, which causes an increase of the luminosity by a factor of $\sim$5-7 on top of the increase of the CR power by a factor of about $\sim$150-550 (depending on the molecular gas distribution). The variation of luminosity for a given CR power or SFR can reach a factor of about 3, depending on the layout of the gas.

\indent Band II (100\mev-100\gev): The luminosity at high gas densities evolves with an index $\delta \sim$1.1-1.2. Compared with band I, the luminosity exhibits larger deviations from this scaling with decreasing gas density, and not only for the small-core case. The scaling index at low gas densities is in the range $\delta \sim$1.4-1.7 depending on the molecular gas distribution. This behaviour is connected with the shift between diffusion-dominated and loss-dominated regimes for the steady-state CR nuclei population (see Sect. \ref{res_gamma}). In the present setup, the transport is not fully loss-dominated over the energy range of particles emitting in the 100\mev-100\gev\ band, even for the highest molecular gas density, and diffusion remains significant (the calorimetric efficiency is at most $\sim$30\%, see Sect. \ref{res_calo}). As a consequence, luminosity still increases modestly with gas density, on top of the increase arising from enhanced star formation.

\indent Band III (100\gev-100\tev): The luminosity evolves with an index $\delta \sim$1.4-1.5 at high gas densities, and with an index $\delta \sim$1.5-1.7 at low gas densities, depending on the molecular gas distribution. The reason for the higher index than in band II is that the steady-state CR nuclei population emitting in band III is closer to the diffusion-dominated side, hence a higher increase of the luminosity for a given increase in gas density. Then, at low gas densities, the effect of an increased fraction of CR power released in the low-density atomic disk is compensated for by the IC contribution, so the decrease in luminosity is more shallow. The variation of luminosity for a given CR power or SFR and depending on the layout of the gas is the strongest over all three bands. This is because particles emitting in this band are significantly affected by diffusion, a process that depends on the distribution of the sources.

\indent The scaling of the \gray\ luminosity for the other star-forming disk galaxy models with $R_{\textrm{max}}=$5 and 20\,kpc follows the same trends. The indices are similar to the values given above for each of the three \gray\ bands, with differences $<0.1$. The variations of the luminosity for a given CR power or SFR and depending on the distribution of the molecular gas are also similar to those given above for $R_{\textrm{max}}=$10\,kpc.

\begin{figure}[!t]
\begin{center}
\includegraphics[width= \columnwidth]{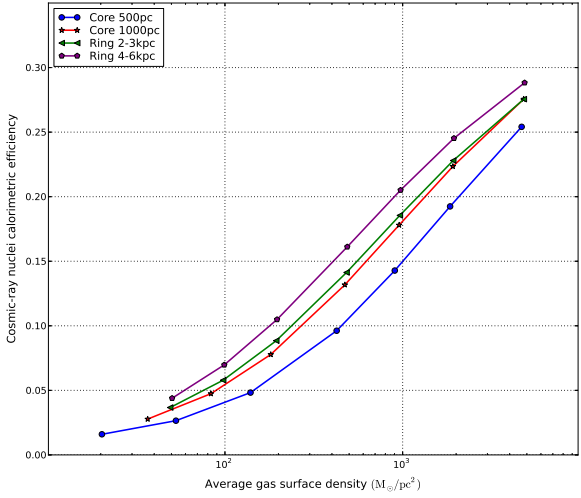}
\caption{Calorimetric efficiency for cosmic-ray nuclei, for a star-forming disk galaxy with $R_{\textrm{max}}=$10\,kpc, 4 profiles of molecular gas distribution, and 7 molecular gas densities. The energy conversion efficiency was computed for hadronic interactions only.}
\label{fig_calo}
\end{center}
\end{figure}
\begin{figure}[!t]
\begin{center}
\includegraphics[width= \columnwidth]{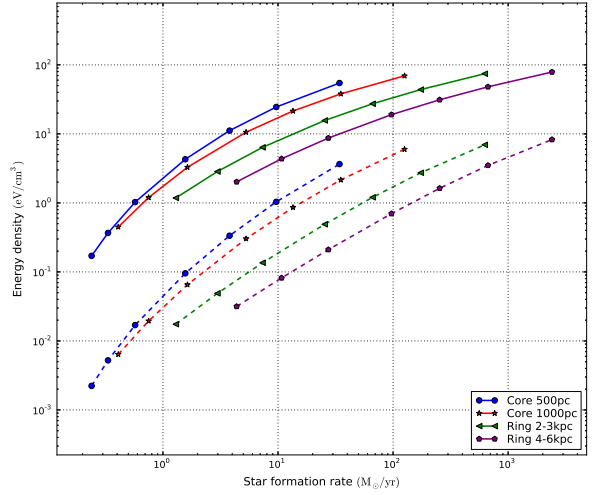}
\caption{Cosmic-ray energy densities for a star-forming disk galaxy with $R_{\textrm{max}}=$10\,kpc, 4 profiles of molecular gas distribution, and 7 molecular gas densities. The energy densities are source-weighted averages, and they are given for particles with energies $\geq$1\gev\ and $\geq$1\tev\ (solid and dashed lines, respectively).}
\label{fig_nrjdens}
\end{center}
\end{figure}

\subsection{Cosmic-ray calorimetry and energy densities}
\label{res_calo}

\indent The anticipation and then detection of star-forming galaxies as HE and VHE \gray\ sources raised the question of how much of the cosmic-ray energy is converted into \grays, and how this fraction would evolve with galactic properties. A more general question is that of the calorimetric efficiency of these galaxies, the fraction of cosmic-ray energy that is lost in the ISM, whatever the process and resulting energy type (radiative or thermal). Nuclei are the dominant CR component, and most of their energy lies above a few 100\,MeV, in a range where hadronic interactions are the main energy-loss process\footnote{Adiabatic losses in a galactic wind are irrelevant since I am interested here in gas-related \gray\ emission and cosmic-ray energy densities in the galactic plane.}. The cosmic-ray calorimetric efficiency can therefore be approximated by the fraction of energy lost to hadronic interactions. Assuming that charged pions are produced at the same rate as neutral pions, the calorimetric efficiency was computed for each galaxy simulated in this work. 

\indent The results are presented in Fig. \ref{fig_calo} for the $R_{\textrm{max}}=$10\,kpc models. The calorimetric efficiencies are given as a function of a source-weighted average gas surface density to illustrate the primary dependence on gas density. Overall, the calorimetric efficiency moves from 2-4\% at the lowest gas densities to 25-30\% at the highest gas densities, which is a rise by one order of magnitude in efficiency for an increase by two orders of magnitude in gas density. These high calorimetric efficiencies are about 10-20 times higher than that inferred for the Milky Way \citep{Strong:2010}. This shows that even at volume-averaged gas densities of about 1000\,H\,cm$^{-3}$, most CR energy flows out of the system. This has at least two reasons: first, CR nuclei with energies of just about 10\gev\ or more can diffuse out of a layer of 100\,pc faster than they lose energy through hadronic interactions; then, a fraction of the lower-energy CRs can also diffuse out without losing too much energy when they are released close to the edges of the dense molecular regions.

\indent Connected with the latter point, another interesting feature of the plot in Fig. \ref{fig_calo} is the vertical scatter in calorimetric efficiency for a given average gas density. This arises from the spatial distribution of gas and cosmic-ray sources. The lowest efficiencies are obtained for the 500\,pc core model, because the latter molecular gas distribution has the highest surface-to-volume ratio. This ratio controls the importance of diffusion (proportional to surface) over energy losses (proportional to volume). This is further demonstrated by the fact that the 1\,kpc core model and the 2-3\,kpc ring model have the same surface-to-volume ratio and nearly identical calorimetric efficiencies, while the 4-6\,kpc ring model has the lowest surface-to-volume ratio of all $R_{\textrm{max}}=$10\,kpc models and the highest calorimetric efficiencies.

\indent The total (nuclei and leptons) CR energy densities resulting from increasing gas densities are shown in Fig. \ref{fig_nrjdens} for the $R_{\textrm{max}}=$10\,kpc models. In the context of pion-decay emission from CR nuclei in the HE and VHE \gray\ bands, the CR energy densities are given as gas-weighted averages for CR energies above 1\gev\ and 1\tev\ (note that the CR energy density above 100\mev\ and 1\gev\ are almost identical). The CR energy density $\geq$1\gev\ is of the order of 0.1-1\,eV\,cm$^{-3}$ at low average gas densities and rises to 50-80\,eV\,cm$^{-3}$ at high average gas densities (depending on the molecular gas distribution). Because of the importance of energy losses in the transport of particles at high gas densities, the increase in CR energy density does not follow the increase in star formation rate. The CR energy density $\geq$1\tev\ is two orders of magnitude lower at low average gas densities, but this difference shrinks to just one order of magnitude at high average gas densities. This occurs because higher-energy particles are closer to a diffusion-dominated regime, so in this range, the CR energy density rise follows more closely the increase in star formation rate (hence cosmic-ray input luminosity).

\indent For comparison, the energy density of radiation and magnetic field rises to $\sim$3300\,eV\,cm$^{-3}$ each for the highest gas densities ($n_{\textrm{H}_2}=500$\,H$_2$\dunit). Cosmic rays are therefore a largely subdominant contribution to the ISM energy density in the densest and most active star-forming galaxies. Their share in the ISM energy density budget increases with decreasing gas density, until it becomes a factor of a few below that of radiation or magnetic field.

\subsection{Impact of model parameters}
\label{res_params}

\indent A subset of models was run with different parameters to assess the extent to which the above discussion depends either on simplifying assumptions of the model or on poorly known aspects of the phenomenon. In the following, these are discussed in turn, with emphasis on the luminosities and their scaling in the three \gray\ bands. The configurations used for these tests are the $R_{\textrm{max}}=$10\,kpc model with the 2-3\,kpc ring distribution and the $R_{\textrm{max}}=$5\,kpc model with the 500\,pc core distribution.

\indent Halo height: The size of the halo for the Milky Way and other galaxies remain a poorly known parameter of the problem. For the Milky Way, a halo size from 4 to 10\,kpc is allowed by measurements of unstable secondary cosmic rays \citep[][depending on propagation model]{Strong:2007}, and \gray\ observations of the outer Galaxy favour a larger than 4\,kpc halo \citep[][among other possibilities]{Ackermann:2012}. A three times larger halo height was therefore tested. This resulted in higher luminosities in all bands since CRs are confined for a longer time and can thus lose a larger part of their energy to radiation. In bands I and II, differences are of the order of 40-50\% at low star formation rates and shrink as molecular gas density increases and transport evolves to an increasingly loss-dominated regime. In band III, differences are of the order of 20-30\% and are almost constant with the star formation rate. This is so because the higher-energy particles emitting in band III are closer to the diffusion-dominated regime over the whole range of gas densities.

\indent Infrared field: The infrared dust emission is a major component in the energy density of the ISM, and dominates starlight for modest molecular gas densities and is as important as the magnetic field for the highest gas densities. It contributes to setting the steady-state lepton populations, therefore I tested cases where it is lower by a factor of $\sim2$. The main effect is on the luminosities in band I, which are lower by about 50\%; the lowest average gas surface densities are affected even more. The reason for this is that decreasing the IR energy density by about 2 reduces the energy losses for leptons, although by less than 2 since synchrotron and Bremsstrahlung are also contributing to the losses. Eventually, the steady-state leptons are more numerous but not to such an extent that it compensates for the reduction in photon number density for inverse-Compton scattering. The luminosity in band II is almost unaffected, which is expected since it is of hadronic origin. The same argument holds for the luminosity in band III except for the lowest average gas surface densities, which are dominated by inverse-Compton emission and thus undergo the same decrease as in band I, by 50\% or more. As mentioned in Sect. \ref{model_isrf}, no radiation transfer calculation was performed and the IR energy density was simply assumed to decrease vertically, away from the galactic plane. Since CRs can spend a significant amount of time in the halo, the actual extent of the IR radiation field may be relevant. A scale height $z_{\textrm{IR}}$ of 20 instead of 2\,kpc was therefore tested. The resulting differences are negligible in all bands, which indicates that regarding inverse-Compton emission everything occurs in the vicinity of the galactic plane .

\indent Diffusion coefficient: The spatial diffusion coefficient was assumed to be the one inferred for the Milky Way from a large set of data interpreted in the framework of a fairly complete GALPROP modelling. But this parameter very likely has some dependence on the actual interstellar conditions in other star-forming galaxies. The tremendous density of star formation in starbursts' cores may result in very strong turbulence such that CRs are expected to be confined more efficiently, diffuse less easily, and lose more energy to radiation. I tested the case of a diffusion coefficient that was ten times lower than the base case value, which yielded \gray\ luminosities higher than those presented previously, although with variations over bands and configurations. In band I, luminosities are almost the same at high densities $n_{\textrm{H}_2} = $500\,H$_2$\dunit, because the transport is loss-dominated in these conditions, but they increasingly differ as the gas density decreases to eventually be a factor 2-3 higher at low densities $n_{\textrm{H}_2} = $5\,H$_2$\dunit. This difference arises mostly from the larger population of secondaries that contribute to the inverse-Compton emission. This is connected to the higher pion-decay luminosity observable in band II. Again, the situation is almost the same at high densities but deviates significantly at low densities, by factors of up to 3-4. Importantly, this low diffusion coefficient moves almost all points above the correlation inferred from {\em Fermi}/LAT observations (see Sect. \ref{res_fermilat}). Last, in band III, the differences are even larger and range between about 2 and 5 because nuclei emitting in this band are more affected by diffusion. With this low diffusion coefficient, inverse-Compton never dominates the emission at very high energies.

\indent Galactic wind: Galactic winds are very common in active star-forming galaxies and can be traced primarily through the expelled gas and dust, but also by the entrained cosmic rays and magnetic fields \citep{Veilleux:2005}. They constitute an energy-independent spatial transport process for CRs, which differs fundamentally from diffusion and is thus likely to alter particle populations and emission spectra. Galactic wind properties are expected to scale with star formation rate surface density, and evidence for this has been observed \citep{McCormick:2013}. Winds are thus expected to be more relevant to star-forming galaxies where most of the star formation rate is confined to small, dense regions. The code used here does not allow the simulation of radially dependent advection, such as a wind emanating only from the inner core of a starburst. I tested the case of a wind with vertical velocity of about 500\kms\ at the level of the galactic plane that accelerates by 100\kms/kpc above the plane. Focusing on the highest gas density and star formation rate cases, which are the most likely to generate such a strong outflow, the wind effect is almost negligible in all \gray\ bands (it is only of a few \%). This can be understood from the time-scale equations given in Sect. \ref{model_timescales}. When considering the time needed to escape a gas layer of 100\,pc thickness, diffusion with the properties assumed here prevails over a 500\kms\ wind over the entire energy range. Time scales become similar (factor of $\sim$2-3) only at energies below a few GeV, but in this range energy losses also occur faster than advection.

\indent Schmidt-Kennicutt relation: The model used in this work relies on the Schmidt-Kennicutt relation for the determination of the star formation rate and its distribution, which in turn affects the CR input luminosity and source profile and the infrared radiation field density. I tested the impact of the index of the Schmidt-Kennicutt relation by varying its value within the uncertainty range and using 1.6 and 1.2 instead of 1.4. Keeping the same normalisation for the dependence of the CR input luminosity on SFR for a given galaxy model with a given gas distribution, a higher (lower) index causes the global SFR to be higher (lower) and more (less) concentrated in the high-density regions, and it increases (decreases) the infrared radiation field density. Apart from the shift in SFR and CR input power, the effects are the following: in band I, the evolution of the luminosity at high gas densities is faster (flatter) in the higher (lower) index case, which arises from variations in the infrared radiation field density; in band II, which is dominated by pion decay, the luminosities are simply increased as a consequence of the higher CR input luminosity, and the scaling of the luminosity with SFR is left essentially unchanged with variations of less than 50\% for a given SFR; in band III, the luminosity evolution with SFR is flattened in the higher index case as a result of a higher contribution of inverse-Compton scattering at low average gas densities (and the opposite for the lower index case). Overall, variations in the index of the Schmidt-Kennicutt relation introduces some scatter about the trend in band II and alters the trend in bands I and III as one moves from high to low SFRs.

\indent The magnetic field model and the impact of its parameters are discussed in Sect. \ref{res_radio} in the context of the constraint set by the observed far-infrared - radio continuum correlation.

\newpage
\begin{figure}[H]
\begin{center}
\includegraphics[width= 8.6cm]{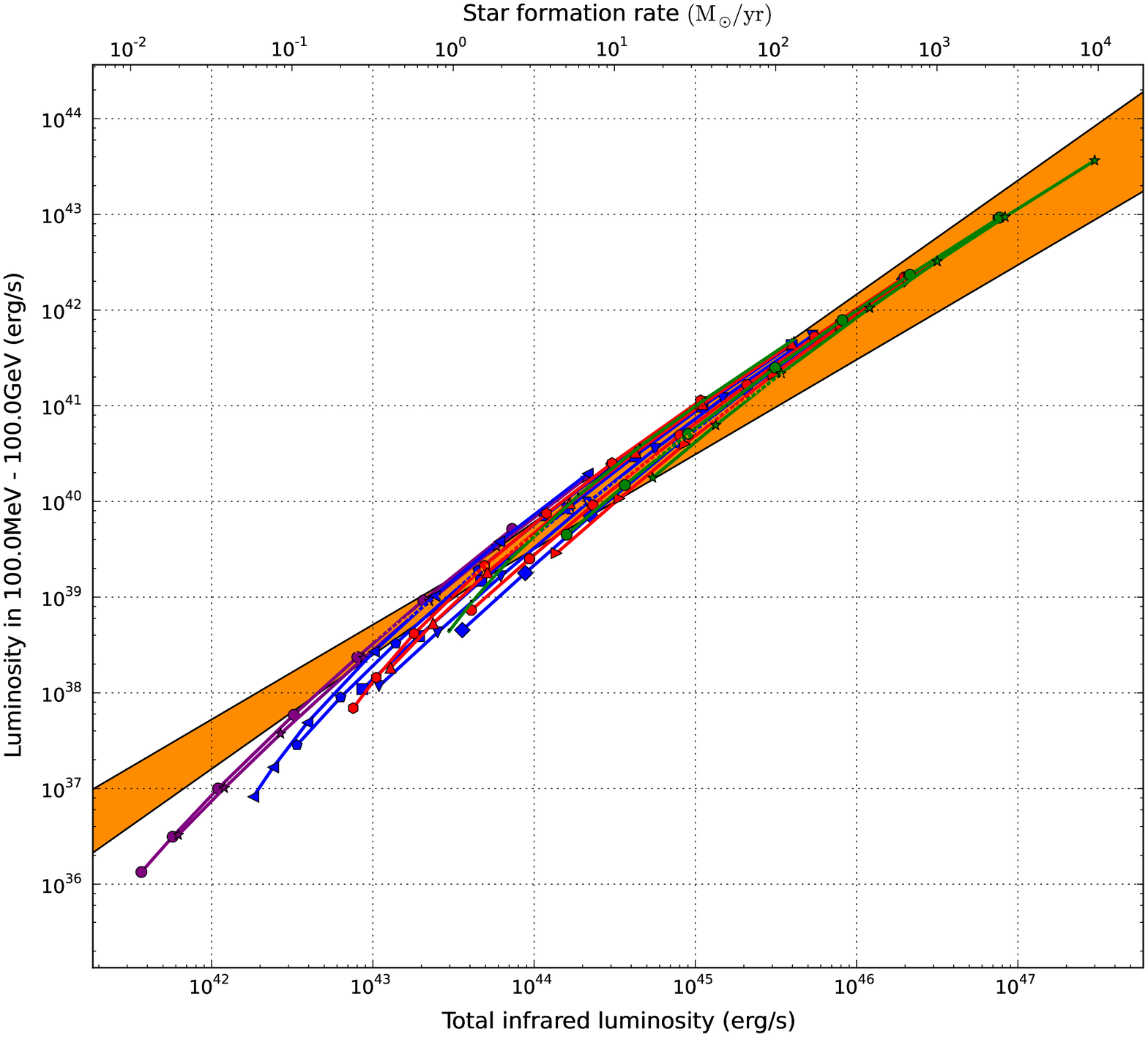}
\includegraphics[width= 8.6cm]{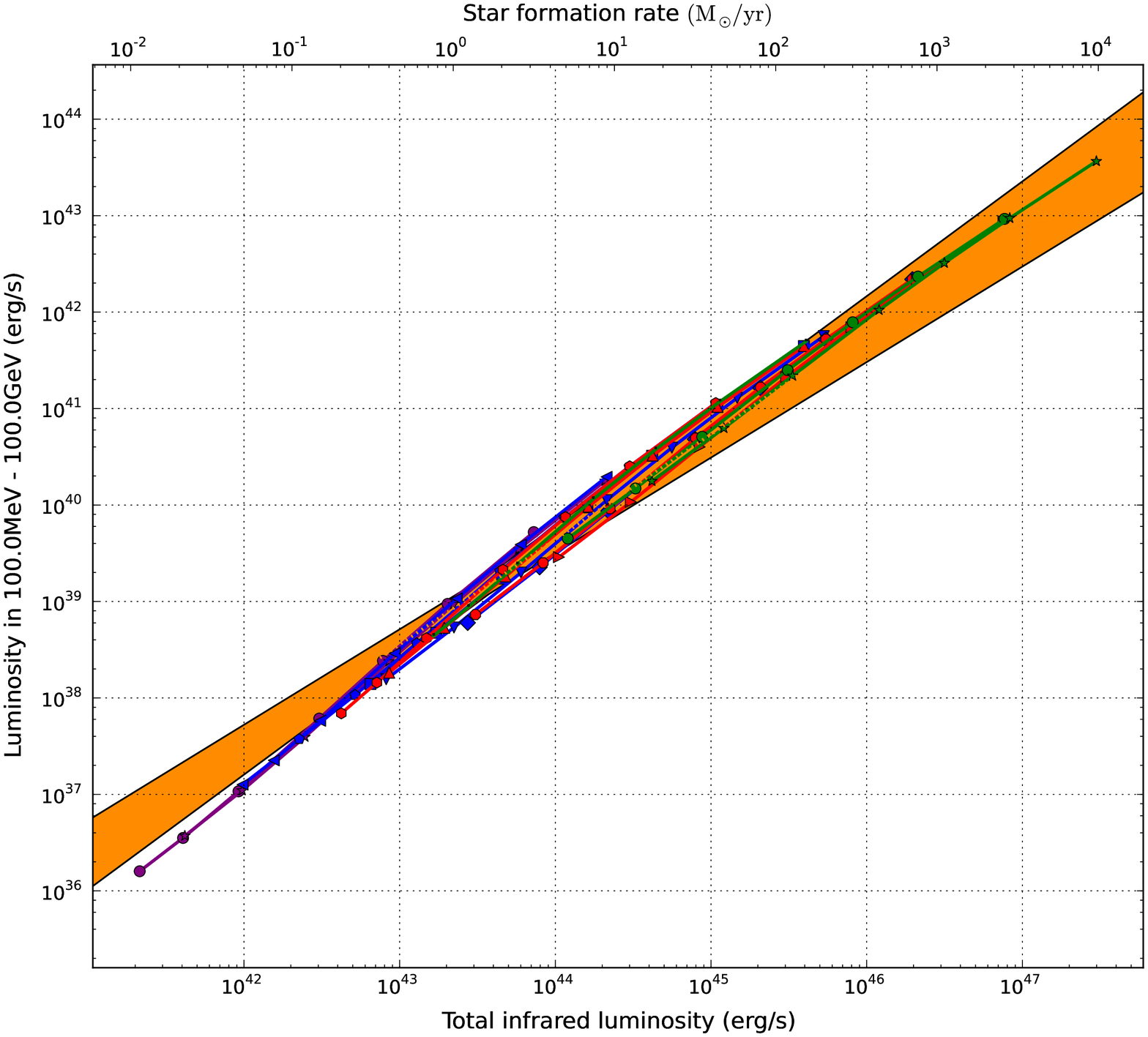}
\includegraphics[width= 8.6cm]{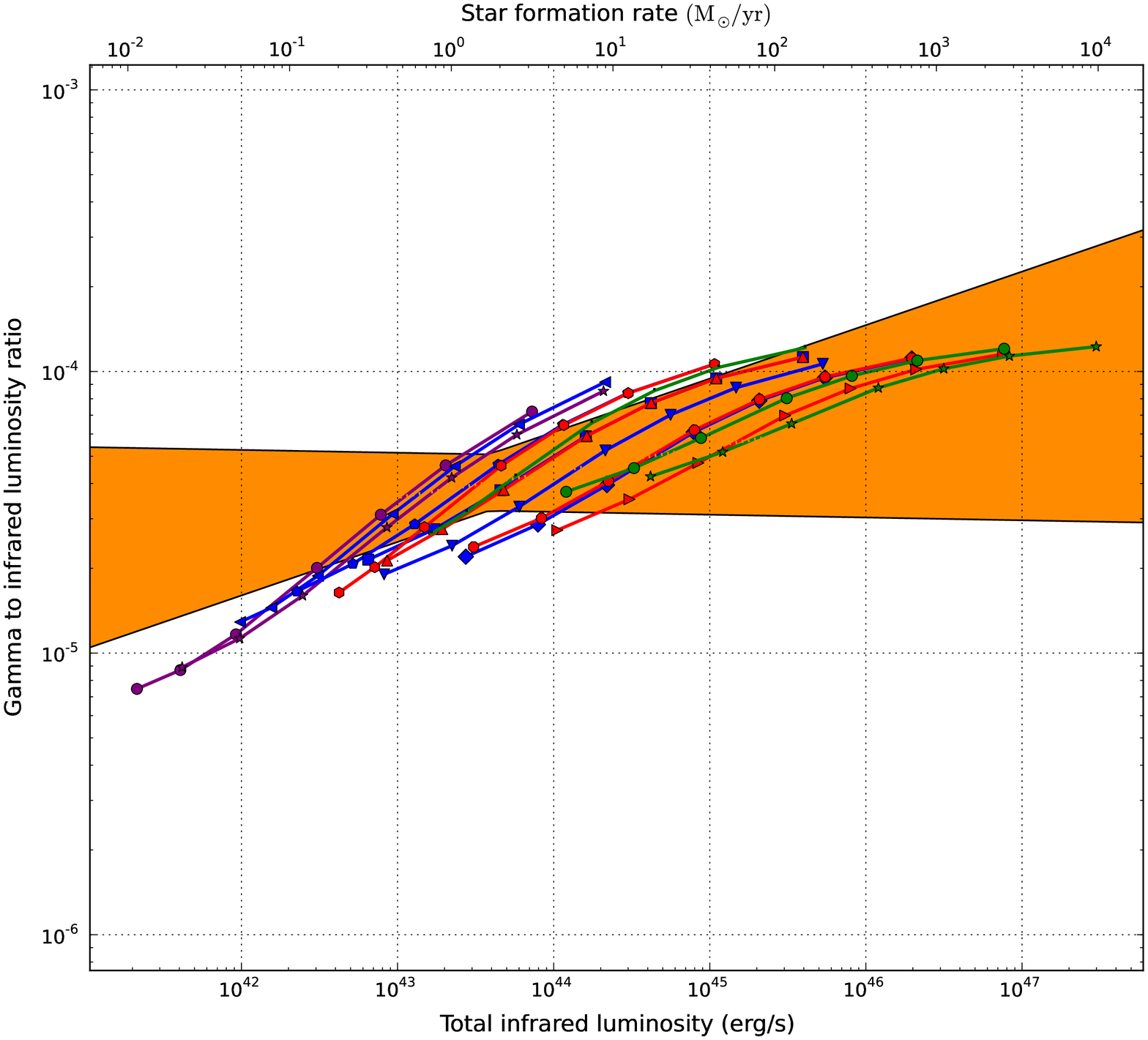}
\caption{Luminosity in the 100\mev-100\gev\ band as a function of total infrared luminosity and star formation rate for the complete sample of models. The orange region is the uncertainty range of the correlation determined experimentally. The top panel shows the result for the basic model, and the middle panel shows the result obtained with larger halos for the small galaxies and using a correction for starlight leakage. The bottom panel shows the ratio of 100\mev-100\gev\ to total infrared luminosities.}
\label{fig_lumgamma_all_vs_fermilat}
\end{center}
\end{figure}

\subsection{Comparison with the Fermi/LAT population study}
\label{res_fermilat}

\indent From observations of about 60 star-forming galaxies using the {\em Fermi}/LAT space telescope, \citet{Ackermann:2012} found evidence for a quasi-linear correlation between \gray\ luminosity and tracers of the star formation activity, although not with high significance (conservative P-values $\leq$ 0.05, for a power-law scaling with index close to 1). Since such a correlation cannot be considered as firmly established, the model introduced above was not tuned to match it. Instead, a generic model was defined, and the corresponding predictions are compared and discussed in the context of the {\em Fermi}/LAT population study. The situation is different for the far-infrared - radio correlation, which is a firmly established empirical correlation with which consistency was sought by defining the magnetic field scaling (see Sect. \ref{res_radio}).

\indent Figure \ref{fig_lumgamma_all_vs_fermilat} shows the luminosity in band II as a function of total infrared luminosity and star formation rate for the complete sample of models. This is compared with the scaling inferred from {\em Fermi}/LAT observations. I used the scaling relationship between $L_{8-1000\mu\textrm{m}}$ and $L_{0.1-100\textrm{GeV}}$, computed using the expectation-maximisation method and excluding galaxies hosting {\em Swift}/BAT-detected AGN, which corresponds to a power-law index of 1.09$\pm$0.10. This does not compare exactly the same things here because the {\em Fermi}/LAT luminosities were derived from power-law fits to the data, whereas the model luminosities correspond to the integration of physical spectra.

\indent The agreement between models and observations is rather good over the nearly 6 orders of magnitude in SFR covered by the models, especially considering that no parameter was tuned to reproduce the inferred correlation. At SFRs of $\sim$1-10\wunit, the models and the correlation uncertainty range clearly overlap. Some models are above or beyond the uncertainty range by factors of up to a few, but they remain in the estimated intrinsic dispersion \citep[not shown here, see][]{Ackermann:2012}. I show below that these outliers can easily be brought to a better agreement with the correlation. At higher SFRs, all relevant models converge with a decreasing scatter towards a scaling that completely agrees with the one inferred from observations (as anticipated from the discussion in Sect. \ref{res_scaling}).

\indent At the lowest densities and SFRs, the luminosities of all models decrease and start deviating from the observed correlation, which contrasts with the fact that the observed correlation appears to hold down to SFRs of 10$^{-2}$\wunit, with objects such as the Small and Large Magellanic Clouds being firmly detected by the {\em Fermi}/LAT. There are, however, two caveats to be considered about the behaviour of the predicted \gray\ luminosity versus SFR at the low end. 
\begin{itemize}
\item The galactic halo of the small galaxies may be larger than assumed here (by keeping a nearly constant aspect ratio of the galactic volume between disk size and halo height).
\item The infrared luminosity may be lower than estimated here for the low gas surface densities, because a fraction of the UV/optical stellar light can escape the galaxy without being converted to infrared.
\end{itemize}
The first effect was assessed using larger halos with $z_{\textrm{max}}$=2\,kpc for the small galaxies with $R_{\textrm{max}}$=2 and 5\,kpc, and the second effect was tested using a correction based on the total gas surface density and an average extinction coefficient for the starlight of $\kappa(1\,\mu\textrm{m})=1.995\,10^{-22}$\,cm$^{2}$\,H$^{-1}$ \citep[][1\,$\mu$m is approximately the peak of the optical radiation field]{Draine:2003}. Combined, they significantly improved the match and reduced the scatter even more, as illustrated in the middle panel of Fig. \ref{fig_lumgamma_all_vs_fermilat}, although a deviation still remains.

\indent For the problem of deviation at low SFRs and/or for small galaxies, one should note that the correlation inferred from {\em Fermi}/LAT observations is not exactly of the same type as the one modelled in this work: only the interstellar \gray\ emission is modelled here, while the total \gray\ emission is used to determine the observed correlation. Most detected objects of the {\em Fermi}/LAT sample are point-like, and for the resolved ones, the Magellanic Clouds, the exact origin of the emission is unclear. For the SMC, it was estimated that the galactic pulsar population may account for a significant fraction of the signal \citep{Abdo:2010e}. A larger sample of detected galaxies with low SFRs would help to clarify the situation. The best prospects here are the detection of the Triangulum Galaxy M33 and a better spatial resolving of the emission from the Magellanic Clouds. At the other end, at high SFRs, doubling the {\em Fermi}/LAT exposure time from five to ten years will lower the constraints but probably not challenge the picture by much, considering that the observed and expected scaling is so close to linear \citep[see][for quantitative estimates of further detections]{Ackermann:2012}.

\indent Overall, the observed correlation can be accounted for to a large extent without major modifications to the global scheme inferred for CRs in the Milky Way. In terms of CR transport, the present results suggest that using a single diffusion coefficient and the assumption that CRs experience large-scale volume-averaged interstellar conditions is enough to account for the population constraints available today \citep[for a recent theoretical investigation of CR sampling of the ISM, see][]{Boettcher:2013}. Interestingly, the same scheme also seems to be able to explain the far-infrared - radio correlation, as illustrated in the following section. The bottom panel of Fig. \ref{fig_lumgamma_all_vs_fermilat} shows the ratio of 100\mev-100\gev\ to total infrared luminosities for the complete sample of models and with the corrections mentioned above. It reveals that the global scaling does not result from a universal relation, but rather from an ensemble of relations that depend to some degree on galactic global properties and are stretched over many orders of magnitude in luminosities.

\indent Last, when exploring possible correlations between the \gray\ luminosity of star-forming galaxies and their global properties both theoretically and experimentally, it was put forward that the luminosity in band II might scale with the product of SFR and total gas mass \citep{Abdo:2009l}. The basic reasoning behind this is that the former quantity traces the input CR power, while the latter quantity traces the amount of target material available for hadronic interactions. Checking that possibility against the models presented here shows that there is indeed a correlation, although with a scatter that is twice as large as that obtained with SFR alone.

\newpage
\begin{figure}[H]
\begin{center}
\includegraphics[width= 8.6cm]{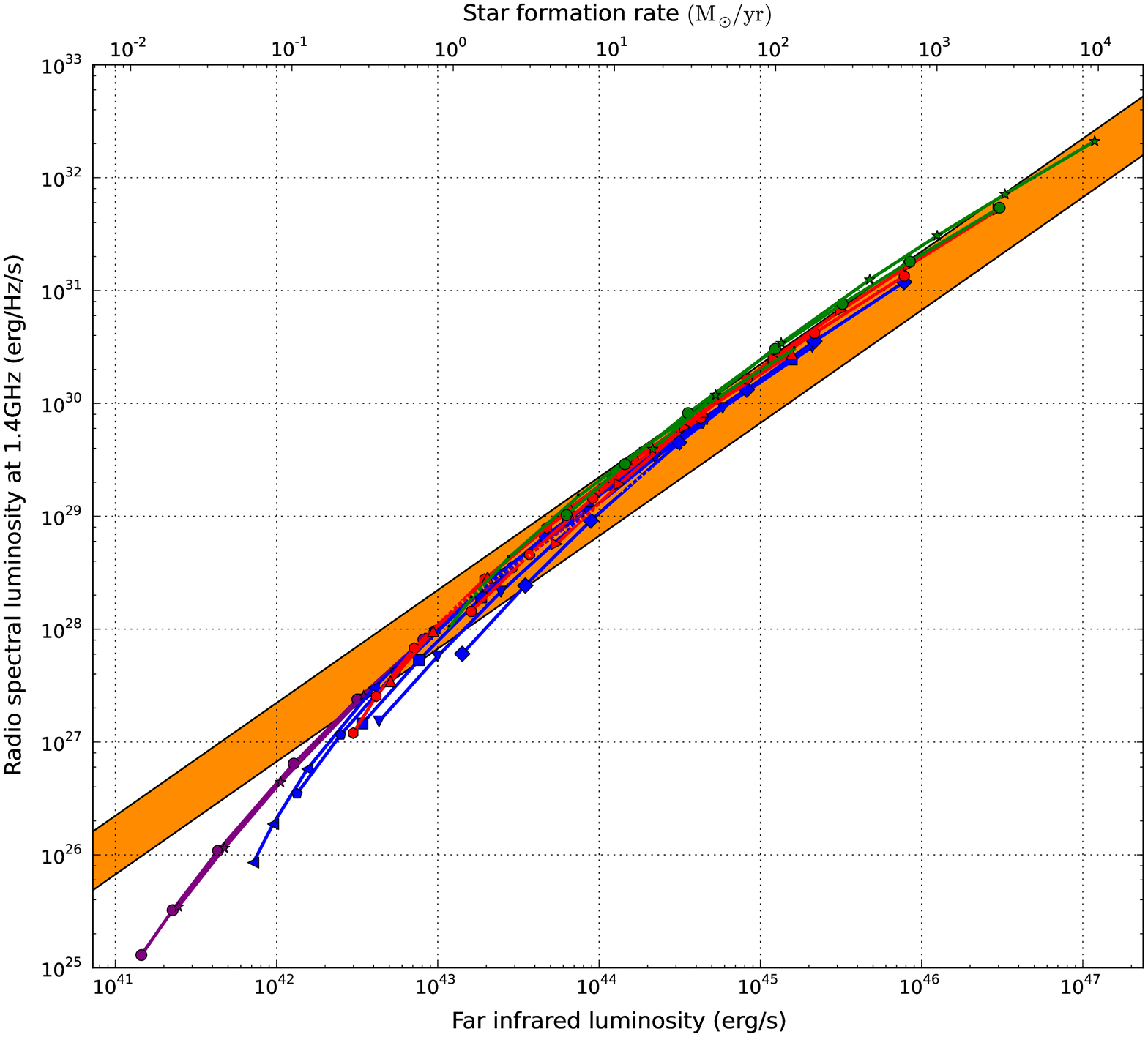}
\includegraphics[width= 8.6cm]{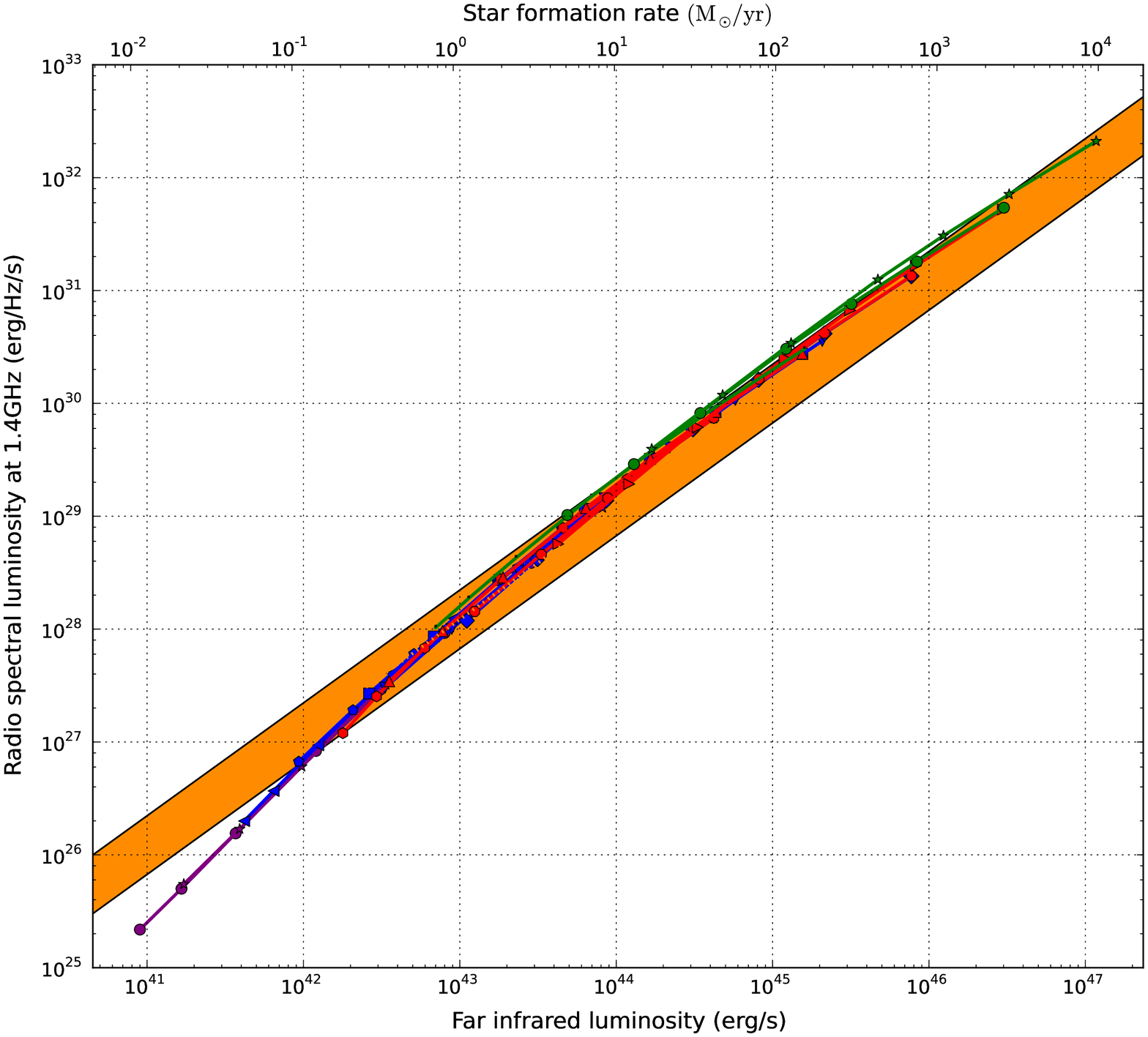}
\includegraphics[width= 8.6cm]{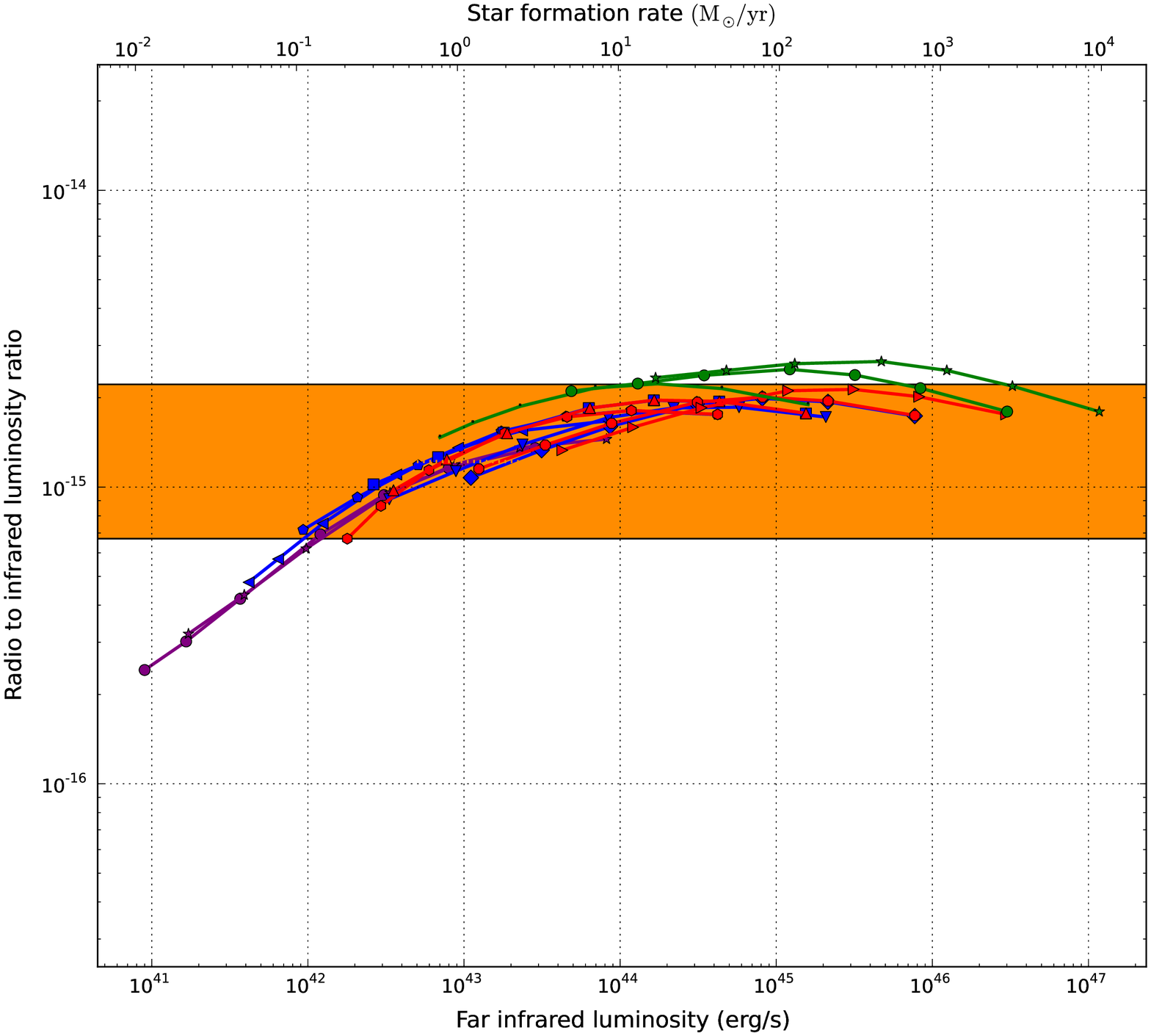}
\caption{Radio luminosity at 1.4\,GHz as a function of far-infrared luminosity for the complete sample of models. The orange region is the uncertainty range of the linear scaling derived from the observed far-infrared - radio correlation. The top panel shows the result for the basic model, and the middle panel shows the result obtained with larger halos for the small galaxies and using a correction for starlight leakage. The bottom panel shows the ratio of 1.4\,GHz to far-infrared luminosities.}
\label{fig_lumradio_all_vs_firradio}
\end{center}
\end{figure}

\subsection{Comparison with the far-infrared-radio correlation}
\label{res_radio}

\indent Before it was possible to probe the \gray\ emission from external star-forming galaxies, mostly thanks to the {\em Fermi}/LAT, the non-thermal radiation from galactic populations of CRs interacting with the ISM had been deeply explored at $\sim$100\,MHz-10\,GHz radio frequencies from the synchrotron emission of CR leptons propagating in the galactic magnetic field. The number of star-forming galaxies detected as a result of this process is orders of magnitude larger than the number of those now accessible through \grays. Population studies can thus lead to much more significant results, and among these is the observed far-infrared - radio correlation. 

\indent Because of the high significance of this correlation, the star-forming galaxy model presented was checked against it before any discussion of the \gray\ counterpart. The definition of the far-infrared/radio flux ratio $q$ was taken from \citet{Helou:1985}: $q=\log_{10}(S_{\textrm{FIR}}/3.75 \times 10^{12}\,\textrm{Hz})-\log_{10}(S_{1.4\,\textrm{GHz}})$, where $S_{\textrm{FIR}}$ (W\,m$^{-2}$) is the FIR flux from 42.5 to 122.5\,$\mu$m and $S_{1.4\,\textrm{GHz}}$ (W\,m$^{-2}$\,Hz$^{-1}$) is the radio flux at 1.4\,GHz. The best-fit value $q=2.34 \pm0.26$ was adopted from \citet{Yun:2001}. Interestingly, estimates for $q$ based on a GALPROP modelling of the Milky Way are in the range 2.26-2.69 \citep[2.45-2.69 for plain diffusion models, see][]{Strong:2010}. The far-infrared and 1.4\,GHz luminosities are readily available for each synthetic galaxy and can thus be compared with the expected values using the measured proportionality factor $q$. The result is shown in the top panel of Fig. \ref{fig_lumradio_all_vs_firradio}.

\indent The index $a$ of the magnetic field model determines the slope of the 1.4\,GHz luminosity evolution for high enough average gas surface densities, while the doublet $(B_0,\Sigma_0)$ sets the normalisation of the 1.4\,GHz luminosities (see Eq. \ref{eq_bfield}). Note that the latter is dependent on the ISRF assumptions because radiation and magnetic field jointly determine the steady-state lepton population. In the star-forming galaxy model used here, index $a$ of the magnetic field prescription was selected so that the predicted radio luminosity matched the far-infrared - radio correlation (although not by a formal fit, but by testing values in the range 0.3-0.8 with steps of 0.1), while the normalisation of the magnetic field was set to reproduce the conditions in the Milky Way (see Sect. \ref{model_calib}). This means that, formally, there was only one degree of freedom to improve the fit to the far-infrared - radio correlation. Despite this, the model results in a fairly good agreement over more than four orders of magnitude, with a remarkably small scatter. Only a few points at high SFRs are marginally out of the uncertainty range of the observed correlation. Deviations appear at the lowest far-infrared luminosities, and especially for the smallest galaxy sizes. This arises because of the low conversion efficiency of CR leptons energy into radio emission, as a result from low magnetic fields (low average gas surface densities) and small confinement volume (small galaxy haloes). 

\indent The same arguments put forward to explain the downturn of the 100\mev-100\gev\ luminosity at low densities and SFRs can be invoked here. In addition, in the context of synchrotron emission, one should note that the magnetic field evolution with gas surface density may differ from the power law assumed here, and may be flatter at low densities for instance. Using larger halos for the small galaxies and a correction for starlight leakage significantly improved the match, as illustrated in the middle panel of Fig. \ref{fig_lumradio_all_vs_firradio}. The resulting scatter in radio luminosity at any given infrared luminosity is relatively small, at most $\pm30$\% and at least twice as small as that of the of 100\mev-100\gev\ luminosity. But although the effects listed above would contribute to bring outliers closer to a single proportionality relation, it is interesting to note that a deviation from a pure linear scaling at low luminosities was observed and discussed \citep{Yun:2001}.

\indent The bottom panel of Fig. \ref{fig_lumradio_all_vs_firradio} shows the ratio of radio-to-infrared luminosities for the complete sample of models and with the corrections mentioned above. Although most models are consistent with the measured value, there is a slight trend in the radio-to-luminosity ratio, corresponding to a variation by a factor of a few, which is concealed by the stretching over 6 orders of magnitude in luminosity. I emphasise here again that only one parameter was adjusted to reproduce the far-infrared - radio correlation, and not through a formal fit. A flatter trend in the radio-to-luminosity ratio may therefore be obtained through a fine-tuning of the model components (magnetic field and ISRF, mainly).

\indent Addressing the latter two problems is beyond the scope of this work. In a forthcoming paper, I will examine in more detail the synchrotron emission as a function of galactic properties as predicted in the framework of the star-forming galaxy model presented here, and connect it with the observed far-infrared - radio correlation.

\section{Conclusion}
\label{conclu}

\indent The interaction of galactic cosmic rays (CRs) with the interstellar medium (ISM) produces \grays\ through hadronic interactions, inverse-Compton scattering, and Bremsstrahlung. This interstellar emission has been observed in the Milky Way for decades and provided a way to indirectly probe CRs outside the solar system, especially the dominant nuclei component. The current generation of high-energy (HE) and very-high-energy (VHE) \gray\ instruments has enabled us to extend this kind of study to several other star-forming galaxies. Five external systems, possibly seven, were recently detected at GeV energies with the \textit{Fermi}/LAT. Combined with upper limits on about sixty other star-forming galaxies, evidence was found for a quasi-linear correlation between \gray\ luminosity and tracers of the star formation activity. This raised the question of the scaling laws that can be expected for the interstellar \gray\ emission as a function of global galactic properties, with the aim of knowing whether this early population study can constrain the origin and transport of CRs.

\indent In the present work, I introduced a model for the non-thermal emissions from steady-state CR populations interacting with the ISM in star-forming disk galaxies. The model is two-dimensional and includes realistic CR source distributions as well as a complete treatment of their transport in the galactic disk and halo (in the diffusion approximation). This contrasts with most previous works, which relied on a one-zone approach and typical time-scale estimates. Prescriptions were used for the gas distribution, the large-scale magnetic field, and the interstellar radiation field. Cosmic-ray-related parameters such as injection rates and spectra or diffusion properties were taken from Milky Way studies. The model is set to reproduce the average interstellar conditions of the Galaxy, its global non-thermal emission, and the far-infrared - radio correlation. A series of star-forming galaxies with different sizes, masses, and gas distributions was simulated to assess the impact of global galactic parameters on the \gray\ output. The link between gas content and star formation was made from the Schmidt-Kennicutt relation. The full set covered almost six orders of magnitude in star formation rate (SFR). The emission was examined over 100\kev-100\tev, with dedicated discussions for instrumental bands: 100\kev-10\mev\ (band I, soft \grays), 100\mev-100\gev\ (band II, HE \grays), and 100\gev-100\tev\ (band III, VHE \grays).

\indent In band I, the emission is dominated by inverse-Compton emission, with Bremsstrahlung contributing at the $\sim20$\% level. With increasing average gas surface density, the luminosity rises more than linearly, with a power-law index on SFR $\sim$1.2-1.3, depending on the gas distribution. This is the result of the growing contribution of secondary leptons. The latter eventually accounts for 70-80\% of the emission.

\indent In band II, the emission is dominated by pion-decay emission, with Bremsstrahlung contributing at the $\sim20$\% level, and inverse-Compton being equally important for small galaxies with low average gas surface density. With increasing average gas surface density, the luminosity rises more than linearly, with a power-law index on SFR initially in the range $\sim$1.4-1.7 and then decreasing to $\sim$1.1-1.2. This behaviour is the result of the progressive shift of the CR nuclei population from a diffusion-dominated regime towards a loss-dominated regime.

\indent In band III, the emission is dominated by inverse-Compton emission for small galaxies with low average gas surface density. For larger and denser systems, pion decay is the main emission mechanism, accounting for more than 95\% of the luminosity for the densest ones. With increasing average gas surface density, the luminosity rises more than linearly, with a power-law index on SFR in the range $\sim$1.4-1.5. The behaviour is the same as in band II, with emitting particles being closer to diffusion-dominated, and with inverse-Compton becoming increasingly important at the low end. 

\indent Without any specific tuning, the model was able to account for the normalisation and trend inferred from the {\em Fermi}/LAT population study over almost the entire range of SFRs tested here. This suggests that the \textit{Fermi}/LAT population study does not call for major modifications of the scheme inferred for CRs in the Milky Way when extrapolated to other systems, probably because uncertainties are still too large. A good match can be obtained with a plain diffusion scheme, a single diffusion coefficient, and the assumption that CRs experience large-scale volume-averaged interstellar conditions. There is no need for a strong dependence of the diffusion properties on interstellar conditions, and no impact of strong galactic winds in the most active star-forming systems. The only requirement in the framework used here is that small galaxies with a disk radius of a few kpc have halos of at least 2\,kpc in half height. There is, however, no universal relation between high-energy \gray\ luminosity and star formation activity, as illustrated by the downturn in \gray\ emission at low SFR values and the scatter introduced by different galactic global properties. Varying the galaxy size, gas distribution, and gas density introduces a variation in the 100\mev-100\gev\ luminosity at a given SFR by up to a factor of 3, while varying the SFR for a given gas layout was found to add another 50\% of scatter. Interestingly, the same model accounted pretty well for the far-infrared - radio correlation over most of the range of galactic properties tested here, with a remarkably small scatter of $\leq 30$\% at any infrared luminosity and a decrease in synchrotron emission at the low end.

\indent Progress on this kind of population study is expected from the doubling of the {\em Fermi}/LAT exposure, especially at low SFRs with the detection of M33 and the resolving of the Magellanic Clouds. An additional test of our understanding of interstellar \gray\ emission from star-forming galaxies may come from the 100\gev-100\tev\ range. In this band, the scaling of luminosity with SFR has a higher non-linearity favourable to distant but more active objects, and the next-generation {\em Cherenkov Telescope Array} with its ten-fold improvement in sensitivity is expected to perform constraining observations.

\begin{acknowledgement}
I acknowledge support from the European Community via contract ERC-StG-200911 for part of this work. I wish to thank Olivier Bern\'e and D\'eborah Paradis for their assistance on interstellar radiation field models, and Andy Strong for his help with the GALPROP package. I also thank Keith Bechtol and Guillaume Dubus for reading and commenting on a draft version of the manuscript.
\end{acknowledgement}

\bibliographystyle{aa}
\bibliography{/Users/pierrickmartin/Documents/MyPapers/biblio/SNRmodels,/Users/pierrickmartin/Documents/MyPapers/biblio/SMC,/Users/pierrickmartin/Documents/MyPapers/biblio/CosmicRaySources,/Users/pierrickmartin/Documents/MyPapers/biblio/CosmicRayTransport,/Users/pierrickmartin/Documents/MyPapers/biblio/GalaxyObservations,/Users/pierrickmartin/Documents/MyPapers/biblio/DataAnalysis,/Users/pierrickmartin/Documents/MyPapers/biblio/Fermi,/Users/pierrickmartin/Documents/MyPapers/biblio/Books,/Users/pierrickmartin/Documents/MyPapers/biblio/SNobservations,/Users/pierrickmartin/Documents/MyPapers/biblio/IMF,/Users/pierrickmartin/Documents/MyPapers/biblio/Positron,/Users/pierrickmartin/Documents/MyPapers/biblio/ISM,/Users/pierrickmartin/Documents/MyPapers/biblio/Cygnus&CygOB2,/Users/pierrickmartin/Documents/MyPapers/biblio/Physics,/Users/pierrickmartin/Documents/MyPapers/biblio/StarFormingGalaxies,/Users/pierrickmartin/Documents/MyPapers/biblio/Starbursts,/Users/pierrickmartin/Documents/MyPapers/biblio/FRC,/Users/pierrickmartin/Documents/MyPapers/biblio/26Al&60Fe}
\newpage

\begin{appendix}
\section{Interstellar radiation field model}

\indent The interstellar radiation field model (ISRF) for the infrared and starlight components is based on \citet{Draine:2007a}. In this work, interstellar dust consists of a mixture of grains and polycyclic aromatic hydrocarbons particles (PAHs), irradiated by the interstellar starlight intensity $I_{\textrm{M83}}(\lambda)$ determined by \citet{Mathis:1983} and scaled by the dimensionless factor $U$ ($U=1$ corresponding to our local Galactic environment). 

\indent For a realistic description of the conditions at the scale of a galaxy, a fraction $(1-\gamma)$ of the total dust mass is exposed to a general diffuse radiation field of intensity $U_{\textrm{min}}$, while a fraction $\gamma$ is heated by a power-law distribution of intensities from $U_{\textrm{min}}$ to a maximum $U_{\textrm{max}}$ and with exponent $\alpha$ \citep[see Eq. 23 in][]{Draine:2007a}. The latter term is assumed to account for the fact that a fraction of the gas and dust will be close to star-forming regions of various sizes and contents. In terms of total dust emission, this is equivalent to having all dust irradiated by a dust-weighted mean starlight intensity $\langle U \rangle$ \citep[see Eq. 33 in][]{Draine:2007a}. The resulting dust emission $I_{\textrm{dust}}(\lambda)$ in the 1-10000\,$\mu$m wavelength band and normalized per hydrogen atom is available for various combinations of the model parameters\footnote{See http://www.astro.princeton.edu/$\sim$draine/. One parameter of the \citet{Draine:2007a} model not mentioned so far is the PAH mass fraction $q_{\textrm{PAH}}$. This parameter is relevant for accurate modelling of infrared spectral energy distributions, but not so much for \gray\ emission. We therefore fixed it at $q_{\textrm{PAH}}=2.50\%$.}.

\indent This ISRF prescription detailed above has several parameters that need to be fixed to provide a generic galaxy model for the study of diffuse \gray\ emission from CRs. The infrared spectral energy distributions of a sample of 65 galaxies of various types were satisfactorily reproduced with $\alpha=2$ and $U_{\textrm{max}}=10^6$ \citep{Draine:2007b}. The same study showed that the diffuse starlight intensity remains close to that of our local environment, with a median value of $U_{\textrm{min}}=1.5$, while the fraction $\gamma$ of the dust mass exposed to higher intensities takes values mostly in the range 1-10\%. I therefore adopted $\alpha=2$, $U_{\textrm{min}}=1.5$, $U_{\textrm{max}}=10^6$, and $\gamma=10\%$ since I focus on actively star-forming galaxies.

\end{appendix}

\end{document}